\documentclass[fleqn,usenatbib]{mnras}

\usepackage{newtxtext,newtxmath}

\usepackage[T1]{fontenc}
\usepackage{ae,aecompl}

\usepackage{graphicx,subcaption}	
\usepackage{float}
\usepackage{caption}
\usepackage{amsmath}	
\usepackage{amssymb}	
\usepackage{amsfonts}

\usepackage{natbib}
\usepackage{color,graphicx} 

\usepackage{wasysym}        

\def\psr{PSR~J1909$-$3744}

\title[J1909$-$3744 revisited]{A revisit of PSR J1909$-$3744 with 15-year high-precision timing}

\author[K.~Liu et al.]{
K.~Liu,$^{1,2}$\thanks{kliu@mpifr-bonn.mpg.de} L.~Guillemot,$^{3,2}$ A.~G.~Istrate,$^{4}$ L.~Shao,$^{5,1,6}$ T.~M.~Tauris,$^{7,8}$ N.~Wex,$^{1}$ J.~Antoniadis,$^{1,9,10}$
\newauthor A.~Chalumeau,$^{3,2}$ I.~Cognard,$^{3,2}$ G.~Desvignes,$^{11,1}$ P.~C.~C.~Freire,$^{1}$ M.~S.~Kehl,$^{1}$ and G.~Theureau$^{3,2,12}$ \\
\\
  $^{1}$Max-Planck-Institut f\"{u}r Radioastronomie, Auf dem H\"{u}gel
  69, D-53121 Bonn, Germany \\
  $^{2}$Station de radioastronomie de Nan\c{c}ay, Observatoire de
  Paris, CNRS/INSU, F-18330 Nan\c{c}ay, France \\
  $^{3}$Laboratoire de Physique et Chimie de l'Environnement et de l'Espace, Universit{\'e} d'Orl{\'e}ans/CNRS, F-45071 Orl{\'e}ans Cedex 02, France \\
  $^{4}$ Department of Astrophysics/IMAPP, Radboud University, PO Box 9010, NL-6500 GL Nijmegen, The Netherlands \\
  $^{5}$Kavli Institute for Astronomy and Astrophysics, Peking University, Beijing 100871, China \\
  $^{6}$National Astronomical Observatories, Chinese Academy of Sciences, Beijing 100012, China \\
  $^{7}$Aarhus Institute of Advanced Studies (AIAS), Aarhus University, H{\o}egh-Guldbergs Gade 6B, DK-8000 Aarhus C, Denmark \\
  $^{8}$Department of Physics and Astronomy, Aarhus University, Ny Munkegade 120, DK-8000 Aarhus C, Denmark \\
  $^{9}$AIFA Argelander Institut f\"{u}r Astronomie, Auf dem H\"{u}gel 71, 53121, Bonn, Germany\\
  $^{10}$Institute of Astrophysics, FORTH, Dept. of Physics, University of Crete, Voutes, University Campus, GR-71003 Heraklion, Greece \\
  $^{11}$LESIA, Observatoire de Paris, Université PSL, CNRS, Sorbonne Université, Université de Paris, 5 Place Jules Janssen, 92195, Meudon, France \\
  $^{12}$Laboratoire Univers et Th\'eories,  Observatoire de Paris, Universit\'{e} Paris-Sciences-et-Lettres, \\
  Centre National de la Recherche Scientifique, Universit\'{e} de Paris, 5 place Jules Janssen, 92195 Meudon, France \\
  }

\date{Accepted XXX. Received YYY; in  original form ZZZ}

\pubyear{2020}

\begin{document}
\label{firstpage}
\pagerange{\pageref{firstpage}--\pageref{lastpage}}
\maketitle
\graphicspath{{plots/}}

\begin{abstract}
We report on a high-precision timing analysis and an astrophysical study of the binary millisecond pulsar, PSR~J1909$-$3744, motivated by the accumulation of data with well improved quality over the past decade. Using 15 years of observations with the Nan\c{c}ay Radio Telescope, we achieve a timing precision of approximately 100\,ns. We verify our timing results by using both broad-band and sub-band template matching methods to create the pulse time-of-arrivals. Compared with previous studies, we improve the measurement precision of secular changes in orbital period and projected semi-major axis. We show that these variations are both dominated by the relative motion between the pulsar system and the solar system barycenter. Additionally, we identified four possible solutions to the ascending node of the pulsar orbit, and measured a precise kinetic distance of the system. Using our timing measurements and published optical observations, we investigate the binary history of this system using the stellar evolution code MESA, and discuss solutions based on detailed WD cooling at the edge of the WD age dichotomy paradigm. We determine the 3-D velocity of the system and show that it has been undergoing a highly eccentric orbit around the centre of our Galaxy. Furthermore, we set up a constraint over dipolar gravitational radiation with the system, which is complementary to previous studies given the mass of the pulsar. We also obtain a new limit on the parameterised post-Newtonian parameter, $\alpha_1<2.1 \times 10^{-5}$ at 95\% confidence level, which is fractionally better than previous best published value and achieved with a more concrete method.
\end{abstract}


\section{Introduction} \label{sec:intro}

Millisecond pulsars (MSPs) are neutron stars (NSs) that were spun up by accretion in a binary system to have a rotational period of $\lesssim30$\,ms \citep{acrs82}. They are well noted for their highly regular rotational behaviour that competes the best atomic clock on Earth over a timescale of decades \citep{mte97,vbc+09}. PSR~J1909$-$3744, a MSP with rotational period of approximately 2.95\,ms, was first discovered in the Swinburne High Latitude Pulsar Survey using the Parkes 64-m Radio Telescope \citep{jbv+03}. Follow-up timing campaigns have achieved a timing precision at the level of a few hundreds nano-seconds \citep[e.g.,][]{dcl+16,rhc+16,abb+18a,krh+20,aab+20a} on a timescale of over ten years, making it one of the most precisely timed pulsars. The timing datasets of PSR~J1909$-$3744 have been utilized for various astrophysical experiments, including placing stringent constraint on nano-Hertz gravitational wave background \citep{src+13,ltm+15,abb+18b}, measuring the mass of main solar system bodies \citep{chm+10,cgl+18} and establishing the first pulsar time standard \citep{hgc+20}.

However, apart from the discovery \citep{jbv+03}, there has not been much work reported that focuses on the astrophysical study of the \psr\; system. The significantly extended timing baseline and the largely improved data quality attributed to a state-of-the-art data recording system in the recent years, strongly motivates a revisit of this pulsar's properties using up-to-date measurements from but not restricted to the radio timing experiment. \psr\; is in a binary system with a $\sim 0.21\;M_\odot$ helium-core white dwarf (He~WD) companion with an orbital period of $P_{\rm b}=1.53\;{\rm days}$. Radio timing analysis has yielded high-precision measurement of its orbital parameters, including a few post-Keplerian parameters from which the masses of the two bodies have been determined precisely \citep[e.g.,][]{dcl+16}. In parallel, spectroscopy and photometry observations of the WD companion in the optical band have been used to infer its temperature, gravity, radius and radial velocity of the system \citep{ant13}. The age of \psr\; is of interest for resolving its formation, evolutionary and kinematic history. While characteristic (spin-down) ages of recycled pulsars are very poor true age estimators \citep[e.g.][]{tau12,tlk12}, the cooling history of their WD companions can be used to determine the age of the systems \citep[e.g.][]{kdk91,vbjj05}.
However, this is a non-trivial exercise given that a cooling age dichotomy exists for extremely low-mass WDs (ELM~WDs) depending on their hydrogen envelope thickness and the associated possibility of undergoing hydrogen shell flashes \citep[e.g.][]{asvp96,althaus2001,itla14,imt+16}. WD masses can also be obtained independently from radio timing data, by using optical measurements in combination with WD cooling models and the mass function of a given binary pulsar \citep[as an example, determining the WD mass from this method also enabled a mass determination of the heavy pulsar PSR~J0348+0432 by][]{afw+13}.


The same as many other NS---WD systems, PSR~J1909$-$3744 is useful for a few types of gravity experiments to constrain alternative theories \citep{wex14}. The asymmetry in the gravitational binding energy of two components, one being a strongly self-gravitating NS while the other being a weakly self-gravitating WD, allows for gravitational dipolar radiative tests in a class of scalar-tensor theories which feature nonperturbative strong-field effects \citep{de93,de96,fwe+12,Shao:2017}. The test depends on the underlying equation of state of NS matters \citep{stob14} and, given the mass of PSR~J1909$-$3744, the measurement of orbital decay could possibly place a better constraint over some equation of states. In addition, the extremely well-measured orbital eccentricity makes PSR~J1909$-$3744 a superb laboratory for testing preferred-frame effects \citep{de92a,sw12,sck+13,sha14, wil14}. The existence of a preferred frame in the Universe would introduce characteristic evolution in the projected semi-major axis and the orbital eccentricity vector.
The absence of abnormal behaviours in them places tight constraints on two (strong-field counterparts of) parameterized post-Newtonian parameters \citep{will18}. The aforementioned tests require high precision timing of the pulsar, preferably measurements of a selection of post-Keplarian parameters and the masses of the system. PSR~J1909$-$3744, as one of the most precisely timed pulsar, has already fulfilled these criteria, and provided improvements in the gravity tests.


The rest of the paper is organized as follows. In Section~\ref{sec:obs} we describe details of the observations and the post-processing of the data. Section~\ref{sec:res} presents the results of the timing analysis, and their application to studying the binary evolution, tracking the galactic motion of the system, and testing alternative theories of gravity. Conclusions are provided in Section~\ref{sec:conclu}.


\section{Observation}
\label{sec:obs}

Regular timing observations of PSR~J1909$-$3744 have been conducted with the Nan\c{c}ay Radio Telescope (NRT) since late-2004. These observations were carried out using the L-band and S-band receivers of the telescope, which have a frequency coverage of 1.1--1.8\,GHz and 1.7--3.5\,GHz, respectively.

Starting from late-2004, the legacy Berkeley-Orl\'eans-Nan\c{c}ay (BON) backend \citep{ct06}, a member of the ASP/GASP coherent dedispersion backend family \citep{dem07}, was used to record the pulsar timing data for nearly ten years, until March 2014. The bulk of the observations with the L-band receiver was conducted initially at a central frequency of 1.4\,GHz and later at 1.6\,GHz after August 2011, while observations with the S-band receiver were mostly performed at a central frequency of 2.0\,GHz. The original bandwidth of these observations of 64\,MHz was increased to 128\,MHz in July 2008. Additional details about the BON dataset used in this paper are given in \cite{dcl+16} which presented the first Data Release from the European Pulsar Timing Array.

The Nan\c{c}ay Ultimate Pulsar Processing Instrument (NUPPI) is a baseband recording system using a Reconfigurable Open Architecture Computing Hardware (ROACH) FPGA board developed by the CASPER group\footnote{http://casper.berkeley.edu/} and Graphics Processing Units (GPUs). It has become the primary pulsar timing backend in operation at Nan\c{c}ay since August 2011 \citep{ctg+13}. The NUPPI dataset included in this paper spans from September 2011 to July 2019. Observations with the NUPPI backend have an integration time ranging from less than 20 to over 80\,min, and a bandwidth of 512\,MHz which is channelised into 128 frequency channels and coherently dedispersed in each. Most of the observations with the L-band receiver were carried out at a central frequency of 1484\,MHz, while those with the S-band receiver are generally centred at 2539\,MHz\footnote{The central frequency of some S-band observations was set to lower frequencies of 1854 and 2154\,MHz, in periods of time when the L-band receiver of Nan\c{c}ay was unavailable.}. The data were polarisation-calibrated with the \textsc{SingleAxis}\footnote{\url{http://psrchive.sourceforge.net/manuals/pac/}} method of \textsc{PSRCHIVE} using observations of a reference noise diode conducted prior to each observation of PSR~J1909$-$3744, to correct for differential phase and amplitude between the two polarisations. Then the data were phase-folded with the pulsar ephemeris from \cite{dcl+16} and visually checked to clean for radio frequency interference (RFI). In the case of L-band observations, the top and bottom 16\,MHz of bandwidth were removed due to the presence of persistent RFI and out-of-band signal reflection in the receiver. We extracted (broad-band) Times-of-Arrivals (TOAs) from the NUPPI data\footnote{The TOAs were measured without fitting for the DM simultaneously, so that the DM modelling is left for the noise analysis stage.} using the Channelised Discrete Fourier Transform (CDFT) algorithm developed in \citet{ldc+14}\footnote{The CDFT approach is integrated into a local branch of \textsc{PSRCHIVE} available at: \url{https://github.com/xuanyuanstar/psrchive_CDFT}.}. In order to form the template profile, we integrated the top three brightest epochs, formed average profiles for each 32-MHz of bandwidth at L-band and 128-MHz of bandwidth at S-band, respectively, and removed the radiometer noise using a wavelet smoothing method. The template-matching procedure was then carried out toward data with the same frequency resolution. For the L-band data, we also divided the entire band into four 128-MHz sub-bands, formed frequency-averaged profiles and calculated the corresponding TOAs using the canonical approach detailed in \citet{tay92a}. These data were later used for comparison purposes as discussed in the next section. In the end, we excluded observations corrupted by calibration issues, intense RFI activity, incidental backend fault, or containing no visible pulsar signal due to interstellar scintillation. Most of the post-processing of the NUPPI data was conducted with the \textsc{PSRCHIVE} software package \citep{hvm04}. In total, the NUPPI dataset includes 405 L-band and 181 S-band observations.


\section{Results}  \label{sec:res}
\subsection{Timing analysis} \label{ssec:timing}

\subsubsection{Dataset and noise properties} \label{sssec:data}

\begin{figure*}
\centering
\hspace*{-2.1cm}\includegraphics[scale=1.1]{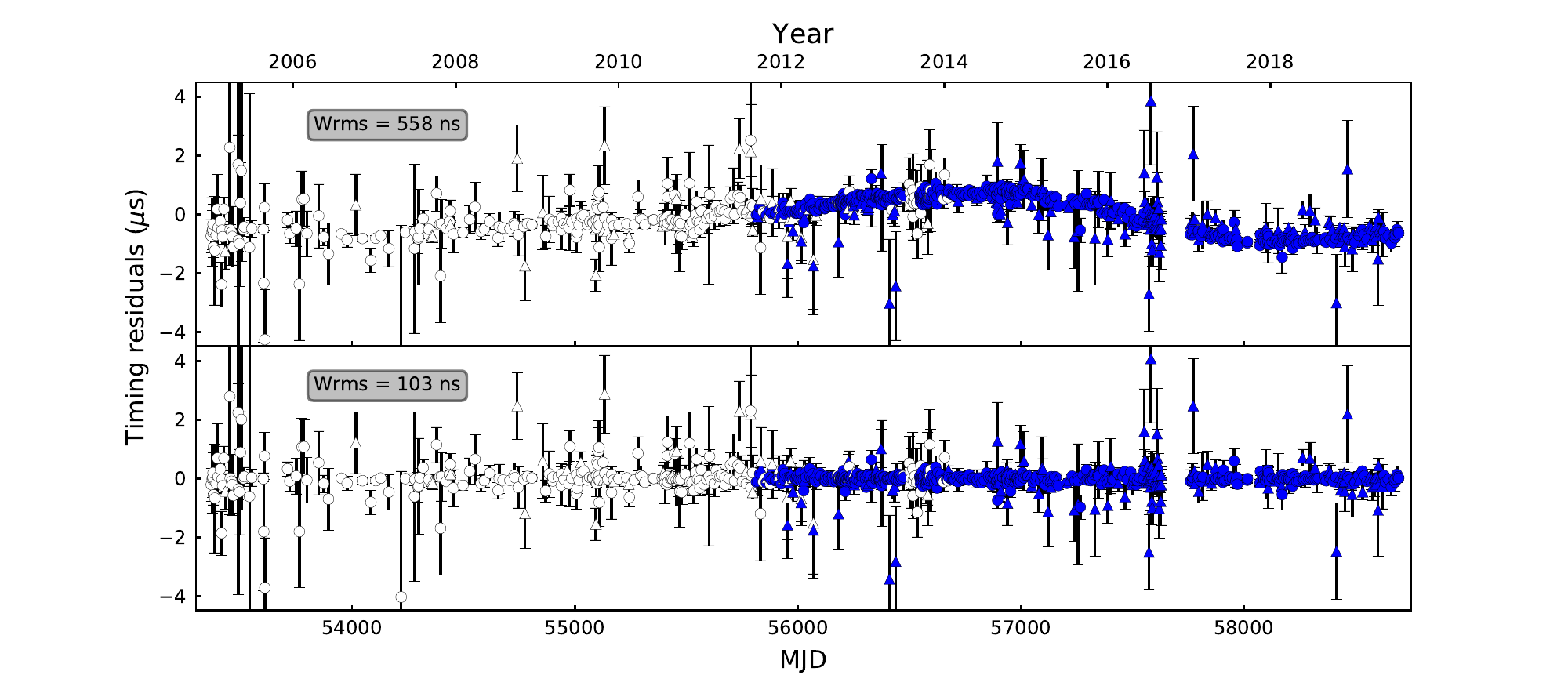}
\caption{Timing residuals of PSR~J1909$-$3744 over a time span of nearly fifteen years. The upper panel shows the residuals before subtracting the stochastic noise model components, and the lower panel shows the residuals after noise subtraction. The circles and triangles represent L-band and S-band observations, respectively. Filled and unfilled symbols stand for NUPPI and BON TOAs, respectively. \label{fig:res}}
\end{figure*}

We combined the BON data presented in \cite{dcl+16} with the NUPPI data collected in the past eight years as described above, which delivered a new dataset with an overall timing baseline of nearly fifteen years. Whenever both BON and NUPPI data were available from the same observation, we kept only the NUPPI data. In total, the dataset contains 846 TOAs, 615 at L-band and 231 at S-band. This gave us an averaged observing cadence of more than one per week. We noted that in an earlier analysis using the NUPPI data of PSR~J1909$-$3744 \citep{ldc+14}, the post-processing of the data was affected by an issue in the \textsc{psrchive} software package\footnote{This was induced by a lack of precision in polyco epoch stored in the header of PSRFITS files. The issue was fixed by an update of \textsc{psrchive} in late 2014.}. Reprocessing the same dataset (which contains 30 epochs from MJD 56545 to 56592) now delivers timing residuals with a weighted rms of approximately 60\,ns, more than a factor of 4 better than the previous value.

We used the \textsc{temponest} software package to perform timing analysis to our dataset. \textsc{temponest} is built based on \textsc{tempo2} \citep{hem06} and \textsc{multinest} \citep{fhb+09} software packages, to explore the parameter space of a non-linear pulsar timing model with Bayesian inference \citep{lah+14}. The fitted parameters included in the timing model are described in Table~\ref{tab:param} (noted as ``\textit{measured parameter}''). To describe the orbital motion of the pulsar in the binary system, we used the ``ELL1'' timing model \citep{lcw+01} as the orbit is known to have a very small eccentricity. Additionally, we included a set of parameters to characterize the noise in the dataset. The white noise parameters are the error factor `EFAC' ($E_{\rm f}$) and an additional error added in quadrature `EQUAD' ($E_{\rm q}$) which relate to the measurement uncertainty ($\sigma_{\rm r}$) of a TOA as:
\begin{equation}
    \sigma=\sqrt{E^2_{\rm q}+E^2_{\rm f}\sigma^2_{\rm r}}.
    \label{eq:whitenoise}
\end{equation}
For each observing system a pair of such parameters was included to capture its white-noise properties to yield a proper weight in the data combination. To describe the long-term red process in the dataset, we included two models to account for the chromatic component caused by dispersion measure (DM) variation and the monochromatic part of the signal, respectively. In both models, the signal is assumed to be a stationary, stochastic process with a power-law spectrum in the form of
\begin{equation}
S(f)=\frac{A^2}{C_{\rm 0}}\left(\frac{f}{f_{\rm r}}\right)^{-\gamma},
\label{eq:rednoise}
\end{equation}
where $S(f)$, $A$, $\gamma$ $f_{\rm r}$ are the power spectral density as a function of frequency $f$, the spectral amplitude, the spectral index and the reference frequency (set to 1\,yr$^{-1}$), respectively. The constant $C_{\rm 0}$ is equal to $1$ for the chromatic term and is $12\pi^2$ for the monochromatic term. The spectrum had a low-frequency cutoff equal to the inverse of the data span (approximately 14.6\,yr), and was sampled with integer multiples of the lowest frequency up to one over fourteen days. To align the data from different systems, we included an arbitrary time offset (known as JUMP) between the reference system and each of the rest systems. Both the timing and noise parameters were simultaneously fitted for during the \textsc{temponest} analysis, while the JUMPs were analytically marginalised. All model parameters were sampled with uniform priors, except for EQUAD and amplitudes of the red processes for which a log-uniform prior was used.

\begin{figure*}
\centering
\begin{subfigure}{.32\textwidth}
\centering
\hspace{-0.3cm}\vspace{-0.1cm}\includegraphics[width=6.0cm,height=4.25cm]{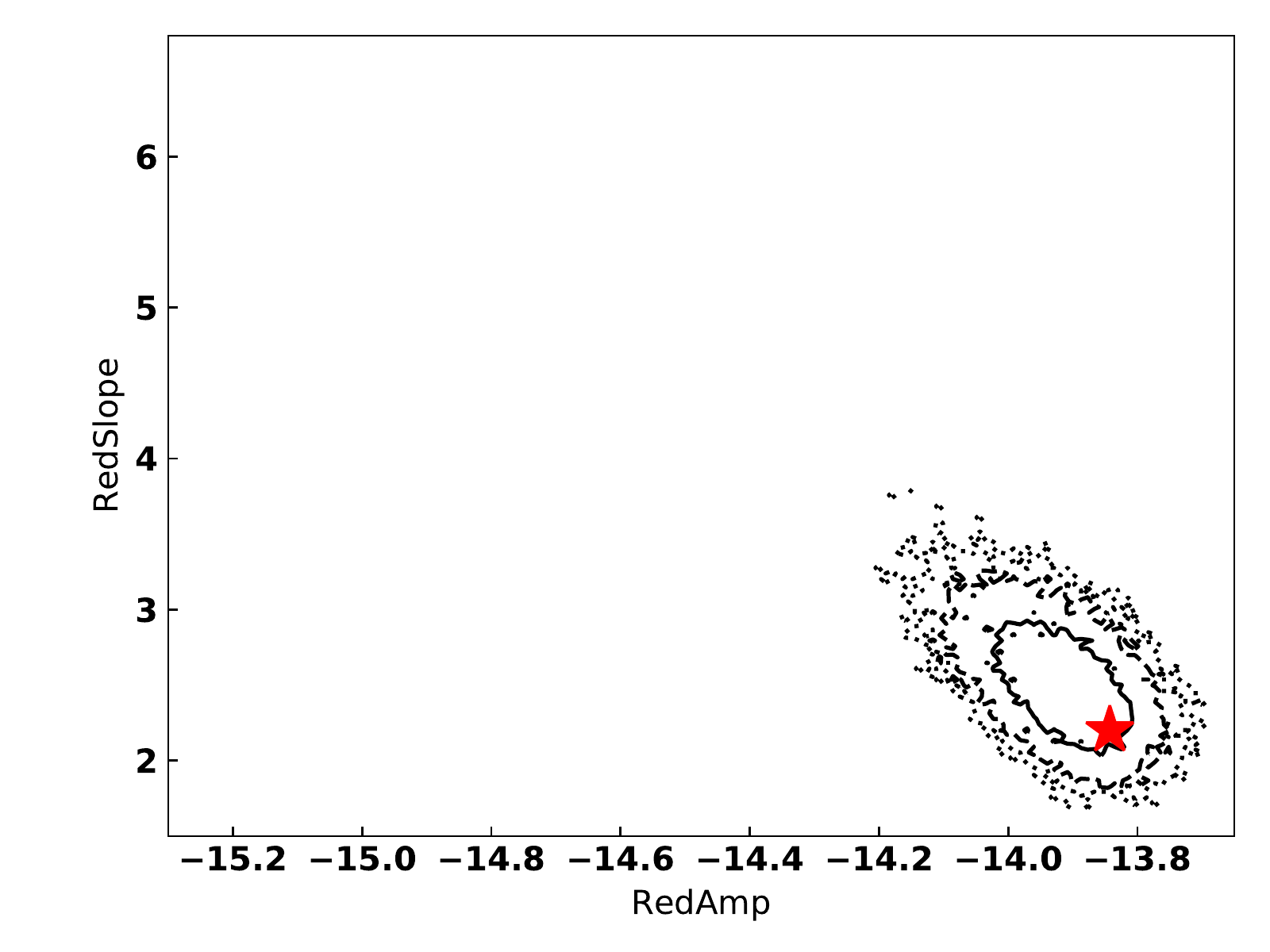}
\includegraphics[scale=0.35]{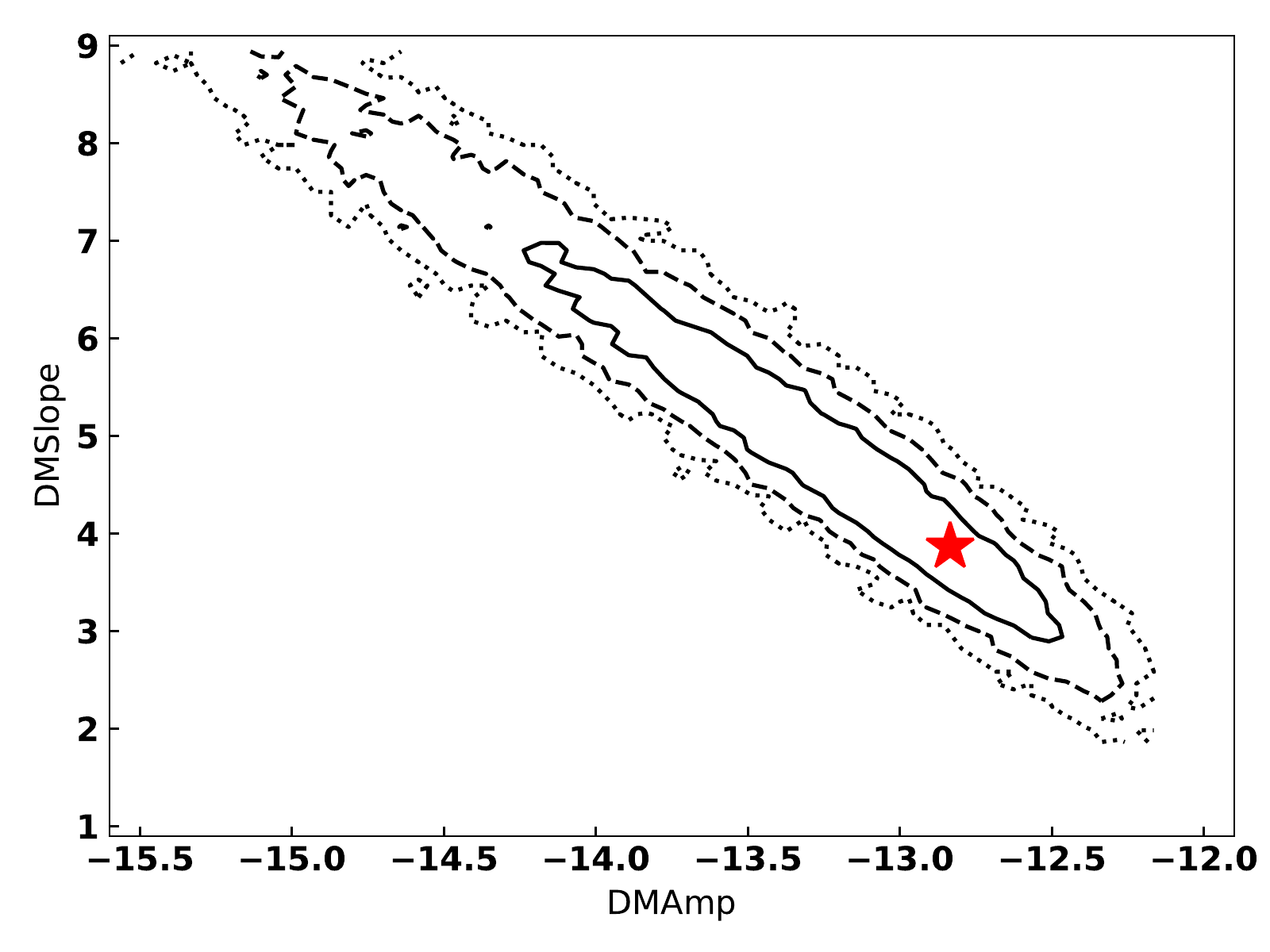}
\caption{Broad-band TOAs, \textsc{temponest}}
\end{subfigure}
\begin{subfigure}{.32\textwidth}
\centering
\includegraphics[scale=0.35]{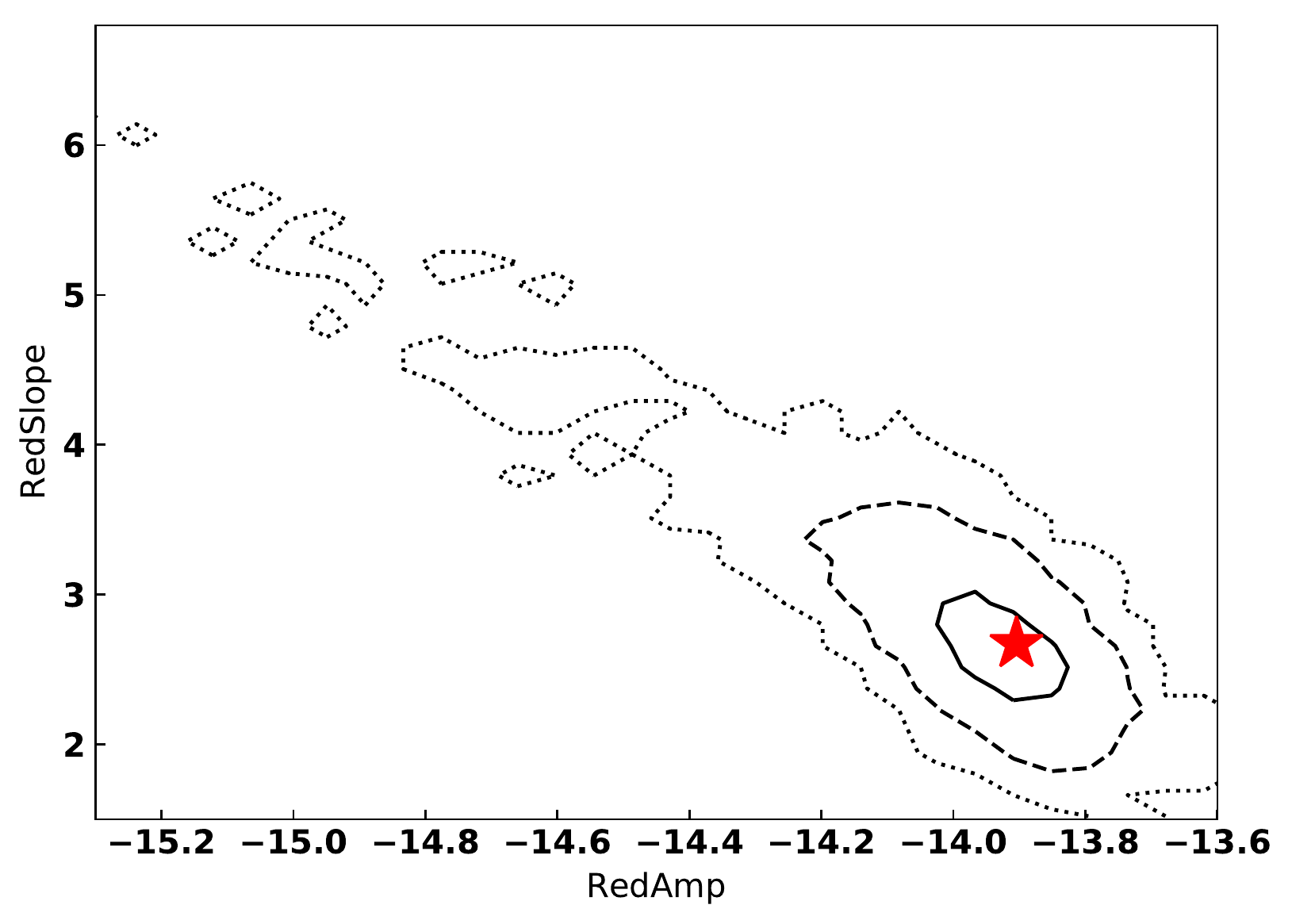}
\includegraphics[scale=0.35]{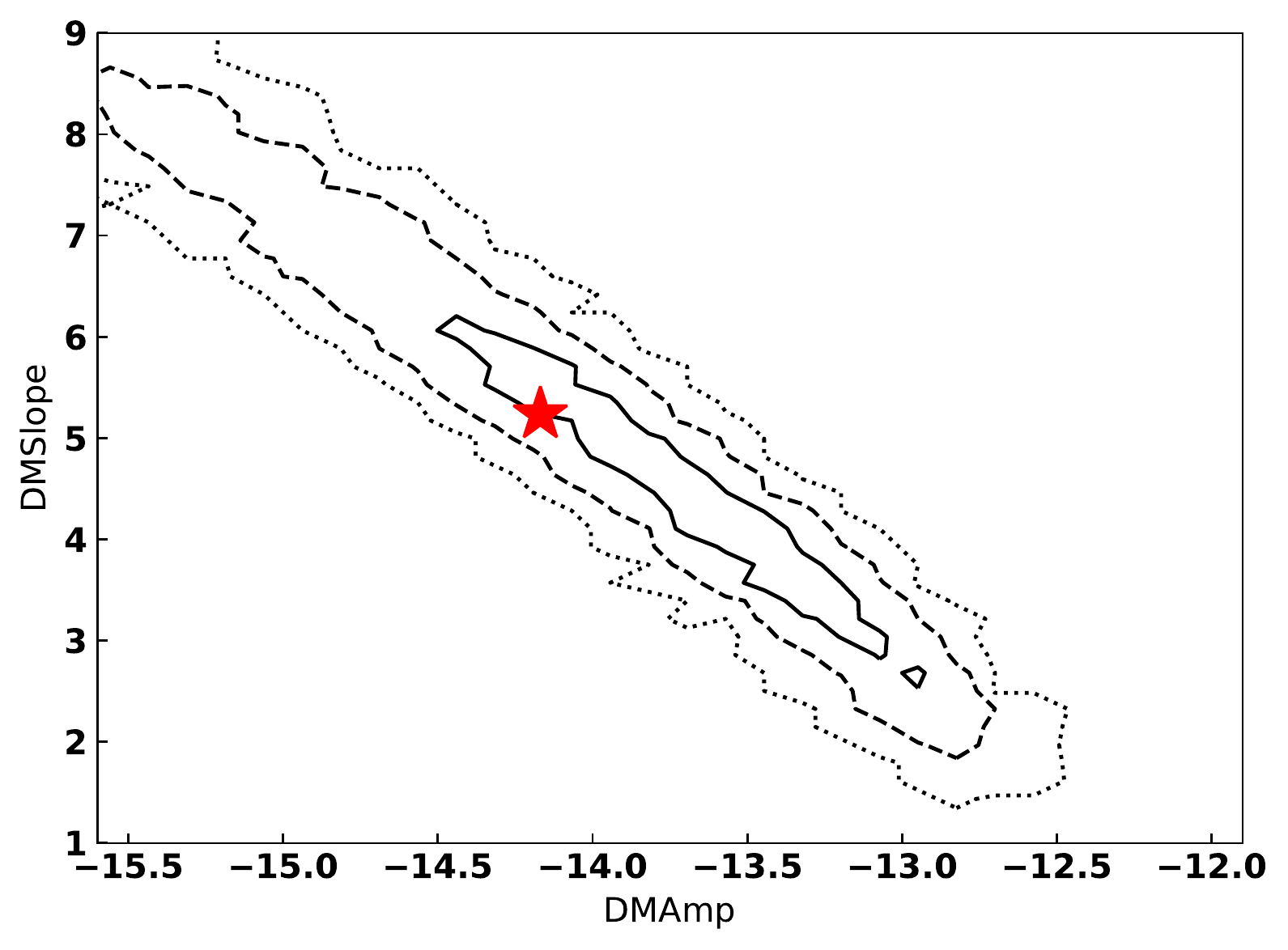}
\caption{Broad-band TOAs, \textsc{enterprise}}
\end{subfigure}
\begin{subfigure}{.32\textwidth}
\centering
\vspace{0.1cm}\includegraphics[scale=0.35]{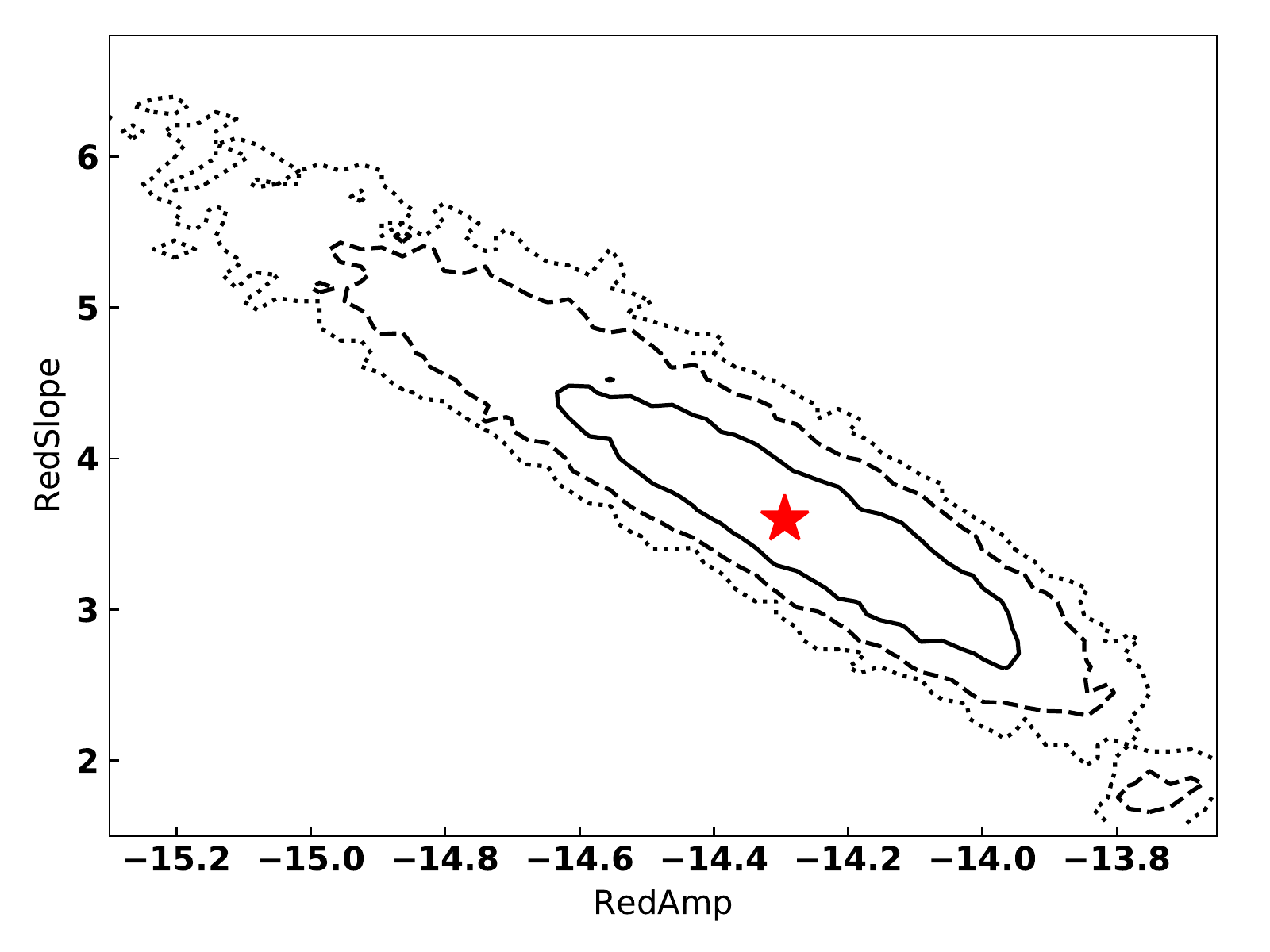}
\includegraphics[scale=0.35]{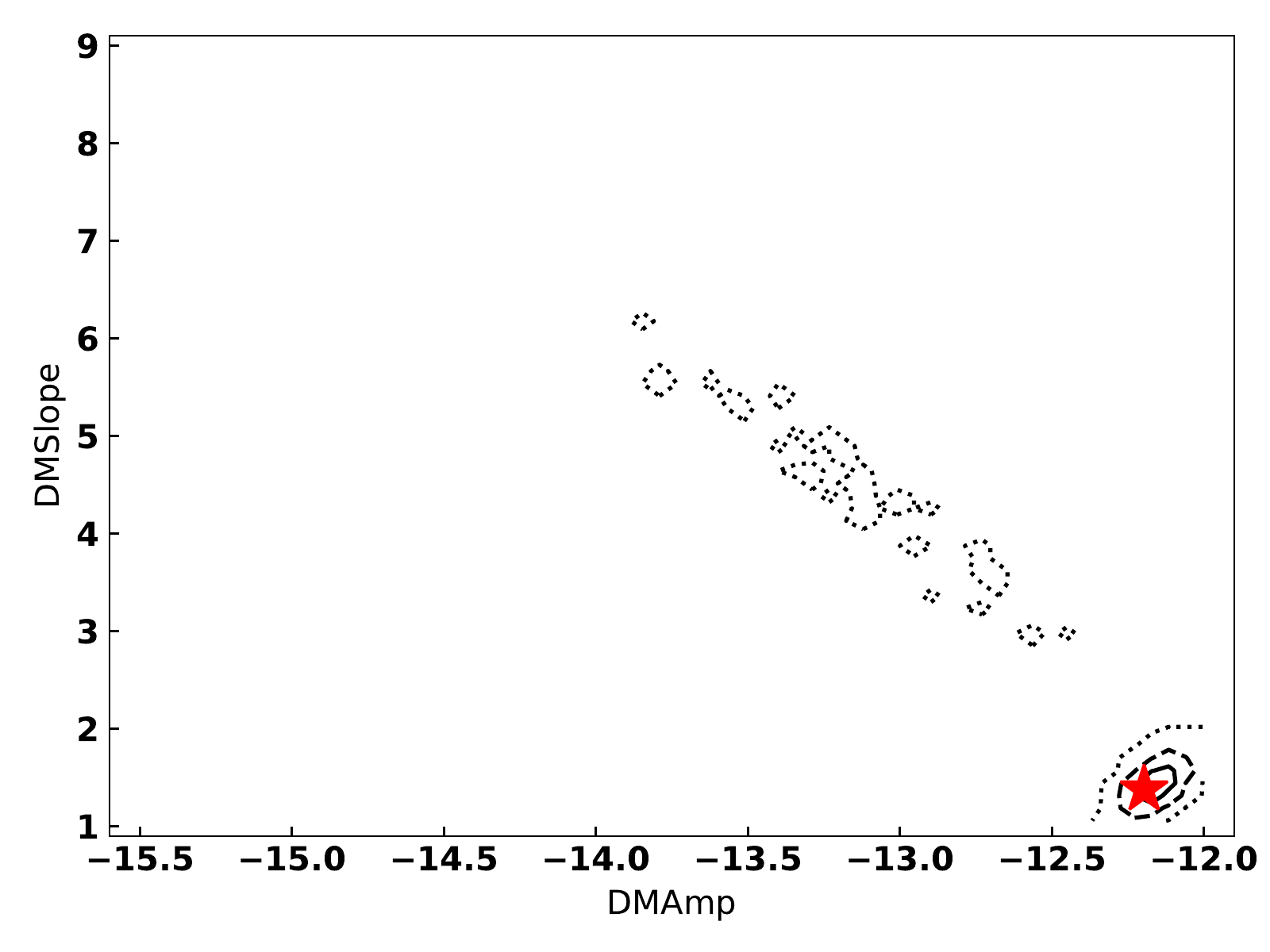}
\caption{Sub-band TOAs, \textsc{temponest}}
\end{subfigure}
\caption{Two dimensional marginalised posterior distributions of the logarithmic amplitude and spectral index of the achromatic red noise and DM noise components, respectively. The solid, dashed and dotted line contours represent 1-$\sigma$, 2-$\sigma$ and 3-$\sigma$ credible regions, respectively. The stars stand for the maximum-likelihood values.}
\label{fig:noisecomp}
\end{figure*}

In Fig.~\ref{fig:res} we present the timing residuals of the dataset obtained from the analysis with \textsc{temponest}, both before and after the subtraction of the red processes in the data. A long-term evolution of the residuals is prominent across the entire time span of the dataset, leading to an weighted rms of over 500\,ns. We note that a similar feature in the PSR~J1909$-$3744 timing residuals was also seen in the recent analysis presented by \cite{abb+18a}. Subtraction of the measured red noise models successfully whitened the residuals, leading to a weighted residual rms of 103\,ns over a timescale of nearly fifteen years. The L-band data from NUPPI itself has a weighted residual rms of 86\,ns. The posterior distribution of the red-noise parameters are presented in Fig.~\ref{fig:noisecomp}, where the logarithmic amplitude and spectral index of the achromatic red noise were measured to be $\log(A_{\rm red})=-13.92^{+0.07}_{-0.07}$ and $\gamma_{\rm red}=2.51^{+0.30}_{-0.27}$, highly consistent with published values by \cite{cll+16} and \cite{abb+18a}.


Timing precision of the brightest MSPs could be limited by the intrinsic variability in the pulsar signal itself which is known as ``jitter noise'' \citep[e.g.,][]{cd85,lvk+11}. To estimate such contribution in our data, we selected three epoch observations when the pulsar signal was the strongest due to interstellar scintillation, and measured the jitter noise following the approach described in \cite{lkl+11}. For L-band, this gave us $\sigma_{\rm J,30min}=14.1\pm1.7$\,ns, where $\sigma_{\rm J,30min}$ is the amount of jitter noise with 30-min integration time. This result is well in agreement with previously published values \citep{sod+14,lcc+16,lma+19}. We did not measure any significant jitter noise in our S-band observation, but obtained a 95\% upper limit of 78\,ns for $\sigma_{\rm J,30min}$ at 2.5\,GHz. The limit is in line with the prediction in \cite{lma+19} that jitter noise of PSR~J1909$-$3744 should be around 10\,ns at this frequency.
To explore the impact of jitter noise on our noise modelling, we incorporated our measurement into the L-band data, i.e., quadratically adding the corresponding jitter noise to the TOA errors before performing the timing analysis with \textsc{temponest}. By doing so, the white noise description here becomes $\sigma=\sqrt{E^2_{\rm q}+E^2_{\rm f}[\sigma^2_{\rm r}+\sigma^2_{\rm J}(\tau)]}$, where $\sigma_{\rm J}$ is calculated based on the integration time ($\tau$) of the observation as $\sigma_{\rm J}=\sigma_{\rm J,30min}/\sqrt{\tau/{\rm 30\,min}}$. In principle, the inclusion of jitter noise in the noise model would be fully covered by EQUAD if the integration times of the observations were identical (i.e., it is then a constant addition to all TOA errors just as EQUAD). This, however, is not the case here because the integration times of our observations range from 20 to 80\,mins, and thus the jitter noise corresponding to each TOA can differ by up to a factor of two.
Using this modified L-band dataset instead in the timing analysis gave a very minimal (well below 10\%) change in EQUAD values of the L-band systems, while the EFAC values remained around unity. This is expected as the jitter noise with a typical 30-min integration time is well below the timing residual rms of the L-band data. In fact, jitter noise is dominant over the radiometer noise only in less than 1\% of our observations.


To further verify our results, we carried out a separate investigation of the noise properties using the \textsc{Enterprise} software package which is also widely used in pulsar timing array data analysis \citep{evtb+19}. We built the noise model consisting of the timing model, white noise and both the chromatic and achromatic red noise components. Those are the same components used in \textsc{temponest}, while the implementation and application are somewhat different. In detail, we performed a fully Bayesian analysis using a parallel tempering Markov Chain Monte Carlo sampler \textsc{PTMCMC} \citep{eh+17}, adopting the same priors as in the \textsc{temponest} analysis but marginalising over the timing parameter errors \citep{vhl+12}. The white noise is characterised by `EFAC' and `EQUAD', with the same description as in \textsc{temponest} (cf. Eq.~\ref{eq:whitenoise}). Both red-noise components were modelled as a stationary Gaussian process with a power-law power spectral density parameterised by an amplitude (at frequency $1/\mathrm{year}$) and a spectral index, the same as in Eq.~\ref{eq:rednoise} except for that here the constant $C_{\rm 0}$ is equal to $12\pi^2$ for both chromatic and monochromatic terms. The inferred posteriors, as shown in  Fig.~\ref{fig:noisecomp}, are largely consistent with the results from \textsc{temponest}, with somewhat more pronounced tail in the distribution for the achromatic red noise distribution. This difference might be attributed to the use of different samplers (\textsc{MULTINEST} vs \textsc{PTMCMC}) in the multidimensional parameter space.

In addition, we noted that the irregularly distributed and unevenly numbered observations between L-band and S-band in the dataset may have an impact on the measurement of DM variation, in particular on separating its contribution to the overall red process from the achromatic component. To examine this potential issue and its potential impact on our timing results, we carried out the same timing analysis to a different version of the dataset where the sub-band TOAs were used for L-band observations with NUPPI. This offers more regular multi-frequency coverage and as shown in Fig.~\ref{fig:noisecomp}, thus provides better constraints on DM model parameters.
The achromatic red noise parameters becomes less constrained, probably due to the drop of the best-estimated value of the amplitude. The spectral index is shifted to a higher value as expected since the effect from an ``imperfect'' DM modelling exhibits as a relatively shallow noise source.
On the other hand, applying the noise model measurements obtained with the sub-band TOAs to subtract the red process in the original dataset resulted in very little difference in the residuals and the overall weighted rms. This suggests that our noise modelling with the broad-band TOAs should still be able to effectively extract the red processes in the data.

\begin{figure}
\centering
\hspace*{-0.5cm}\includegraphics[scale=0.6]{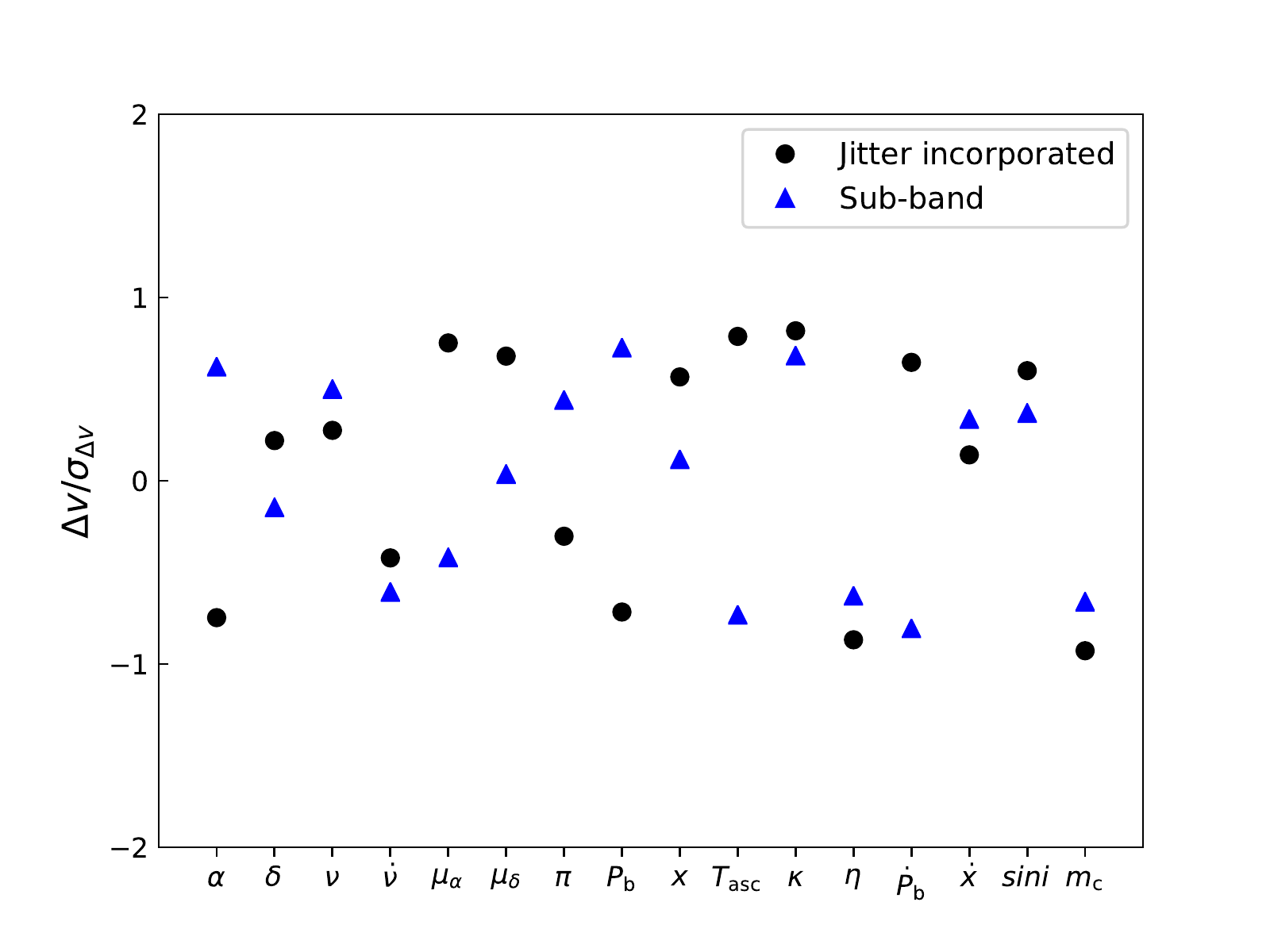}
\caption{Differences (normalised by their errors) in measured timing parameters, from those obtained with the original dataset which contains broad-band TOAs without jitter noise incorporated in their uncertainties. \label{fig:timingcomp}}
\end{figure}

It is worth noting that the experimental analysis on incorporating jitter noise measurement in the TOAs and using sub-band TOAs in the dataset have both yielded highly consistent measurement in all timing parameters, compared with those from using the broad-band TOAs. This is demonstrated in Fig.~\ref{fig:timingcomp}, where the differences in measured timing parameters are shown to be all consistent with zero within 1-$\sigma$ confidence interval. We note that in \cite{aab+20a} and \cite{aab+20b} the analysis using broad-band and sub-band TOAs has also led to highly consistent timing results. Thus, for the analysis in the rest of the paper we will simply quote the timing results obtained from the broad-band dataset. A more detailed noise analysis using different forms of datasets will be presented in an upcoming paper (Chalumeau et al. in prep.).

\subsubsection{Astrophysical measurements} \label{sssec:astro measure}

\begin{figure*}
\centering
\hspace*{-2cm}\includegraphics[scale=1.35]{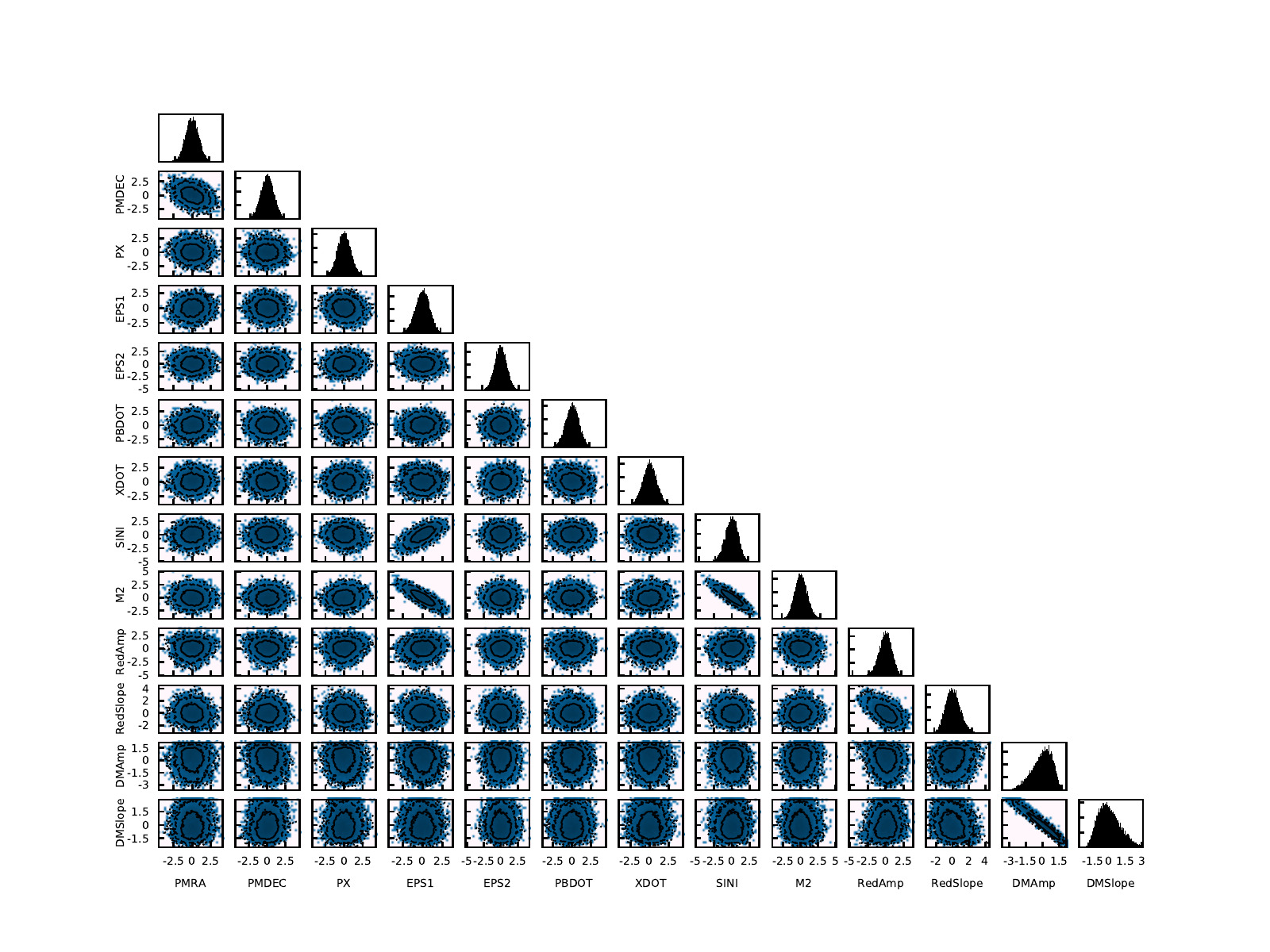}
\caption{Two dimensional marginalised posterior distribution for a subset of the timing and noise parameters: proper motion in right ascension (PMRA) and declination (PMDEC), annual parallax (PX), orbital eccentricity in component $\hat{x}$ (EPS1) and $\hat{y}$ (EPS2), derivative of orbital period (PBDOT), derivative of orbital projected semi-major axis (XDOT), shape of the Shapiro delay (SINI), range of the Shapiro delay (M2), logarithmic amplitude (RedAmp) and spectral index (RedSlope) of the achromatic red noise and those of the DM noise component (DMAmp, DMSlope). The distributions are with the maximum-likelihood values (presented in Table~\ref{tab:param} and Fig.~\ref{fig:noisecomp}) subtracted and normalised by their standard deviations. \label{fig:corner}}
\end{figure*}

\begin{figure}
\includegraphics[scale=0.4]{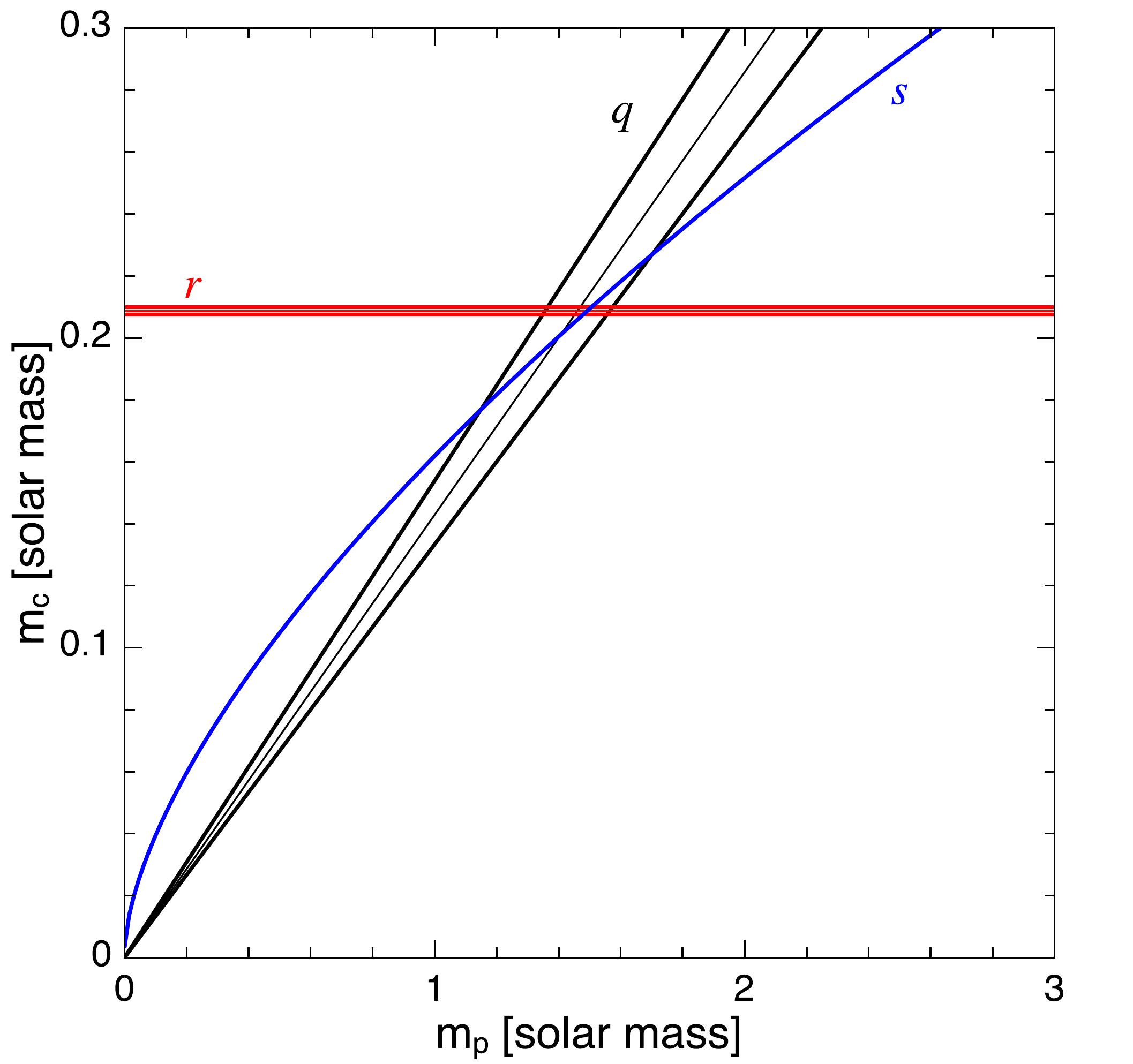}
\caption{Mass--mass diagram of the \psr{} system which shows the constraints from the Shapiro delay parameters, $r$ and $s$, in our timing analysis, and from mass ratio, $q$, measurement in \protect\cite{ant13} based on optical observations of the WD. \label{fig:m-m}}
\end{figure}

The results of the timing parameters from the analysis are summarised in Table~\ref{tab:param}, where the uncertainties of the measurements are 1-$\sigma$ Bayesian credible interval obtained from one-dimensional marginalised posterior distribution. The posterior distributions of a subset of the timing and noise parameters from the \textsc{temponest} analysis can be found in Fig.~\ref{fig:corner}. In Table~\ref{tab:paramcomp} we compared a subset of timing parameters from our analysis with a few recent publications. It can be seen that all of our measurements have a broad consistency with previous analysis, and that our work has led to a clear improvement in measurement precision of $\dot{P}_{\rm b}$ and $\dot{x}$. We note that \cite{aab+20a} reported a covariance between $\dot{x}$ and the red-noise terms. This, however, is not obvious from the posteriors shown in Fig.~\ref{fig:corner}, possibly because the longer timing baseline here helps to mitigate the degeneracy between $\dot{x}$ and the red-noise parameters. Using the measurements of the Shapiro delay parameters, $s$ and $r$, the mass of the pulsar and the WD are derived to be $m_{\rm p} = 1.492\pm0.014$\,$M_{\odot}$ and $m_{\rm c} = 0.209\pm0.001$\,$M_{\odot}$, respectively, as demonstrated in Fig.~\ref{fig:m-m}. These new values resolves the slight inconsistency reported in \cite{dcl+16} from other works. Fig.~\ref{fig:m-m} also shows the independent constraint on the mass ratio ($q \equiv m_{\rm p}/m_{\rm c} = 7.0 \pm 0.5$) reported in \cite{ant13} based on optical observations of the WD, which is consistent with the values determined from Shapiro delay, albeit with a larger uncertainty.

The orbital eccentricity can be calculated as $e = \sqrt{\kappa^2+\eta^2} = (1.15 \pm 0.07) \times 10^{-7}$, which still makes \psr\; the most circular system in all binary pulsars known by far \citep{mhth05}. Using the mass measurements, the relativistic periastron advance of the system is calculated to be 0.14\,deg/yr \citep{lk05}. Accordingly, the eccentricity vector has rotated by only 2\,deg within the fifteen years of our timing baseline, which, if unmodelled, would cause an apparent variation in the $\hat{x}$, $\hat{y}$ components of the eccentricity vector by less than $10^{-10}$. This is well below their measurement precision shown in Table~\ref{tab:param}. Thus, the periastron advance is still not measurable with our dataset.

\begin{table}
\centering \caption{Measured and derived timing parameters of PSR~J1909$-$3744.
\label{tab:param}}
\begin{tabular}[c]{ll}
\hline
\hline
Parameter  & Value \\
\hline
MJD range &  53368$-$58693  \\
Number of TOAs & 846  \\
Timing residual rms ($\mu$s) &  0.103 \\
Reference epoch (MJD) & 55000 \\
\hline
\textit{Measured parameter} & \\
Right ascension, $\alpha$ (J2000) & 19:09:47.4335812(6) \\
Declination, $\delta$ (J2000) & $-$37:44:14.51566(2) \\
Proper motion in $\alpha$, $\mu_{\rm\alpha}$ (mas\,yr$^{-1}$) & $-$9.512(1) \\
Proper motion in $\delta$, $\mu_{\rm\delta}$ (mas\,yr$^{-1}$) & $-$35.782(5) \\
Parallax, $\pi$ (mas) & 0.861(13) \\
Spin frequency, $\nu$ (Hz) & 339.315687218483(1) \\
Spin frequency derivative, $\dot{\nu}$ & $-1.614795(7)\times10^{-15}$ \\
DM (cm$^{-3}$~pc) & 10.3928(3) \\
DM1 (cm$^{-3}$~pc~yr$^{-1}$) & $-$0.00035(5) \\
DM2 (cm$^{-3}$~pc~yr$^{-2}$) & $2.2(7)\times10^{-5}$ \\
Orbital period, $P_{\rm b}$ (d) & 1.533449474305(5) \\
Epoch of ascending node (MJD), $T_{\rm asc}$ & 53113.950742009(5) \\
Projected semi-major axis, $x$ (s) & 1.89799111(3) \\
$\hat{x}$ component of the eccentricity, $\kappa$ &  $4.68(98)\times10^{-8}$\\
$\hat{y}$ component of the eccentricity, $\eta$ & $-1.05(5)\times10^{-7}$\\
Orbital period derivative, $\dot{P}_{\rm b}$ & $5.1087(13)\times10^{-13}$ \\
Derivative of $x$, $\dot{x}$ & $-2.61(55)\times10^{-16}$ \\
Shape of Shapiro delay, $s$ & 0.998005(65) \\
Range of Shapiro delay, $r$ ($\mu$s) & 1.029(5) \\
\hline
\textit{Derived parameter (assuming GR)} & ~~ \\
Galactic longitude, $l$ (deg) & 359.7 \\
Galactic latitude, $b$ (deg) &$-$19.6 \\
Longitude of periastron, $\omega$ (deg) & 156(5) \\
Orbital eccentricity, $e$ & $1.15(7)\times10^{-7}$ \\
Pulsar mass, $m_{\rm p}$ ($M_{\odot}$) & 1.492(14) \\
Companion mass, $m_{\rm c}$ ($M_{\odot}$) & 0.209(1) \\
Parallax distance, $d_{\pi}$ (kpc) & 1.16(2) \\
kinematic distance, $d_{\rm k}$ (kpc) & 1.158(3) \\
Spin period, $P$ (ms) & 2.94710806976663(1) \\
Spin period derivative, $\dot{P}$ ($\times10^{-21}$) & 14.02521(6)\\
$\dot{P}_{\rm Gal}$ ($\times10^{-21}$) & 0.0587(2) \\
$\dot{P}_{\rm Shk}$ ($\times10^{-21}$) & 11.36(3) \\
$\dot{P}_{\rm Int}$ ($\times10^{-21}$) & 2.60(3) \\
Characteristic age, $\tau_{\rm c}$ (Gyr) & 18.0 \\
Surface magnetic field, $B$ (G) & $8.9\times10^7$ \\
\hline
\hline
\end{tabular}
\end{table}

\begin{table*}
    \centering
     \caption{Comparison of a selection of timing parameters measured in this work with previous publications. The values of orbital eccentricity $e$ were calculated from two eccentricity vectors $\kappa$, $\eta$, except for the case of \protect\cite{rhc+16} where it was directed fitted for. \label{tab:paramcomp}}
    \begin{tabular}{ccccccc}
    \hline
    \hline
    &$\pi$ (mas) &$\dot{P}_{\rm b}$ &$\dot{x}$ & $\sin{i}$ & $m_{\rm c}$ ($M_{\odot}$) &$e$  \\
\hline
\cite{dcl+16}  & 0.87(2)  &$5.03(5)\times10^{-13}$  & $-0.6(17)\times10^{-16}$ &0.99771(13) &0.213(2) &$1.22(11)\times10^{-7}$ \\
\cite{rhc+16}  & 0.81(3)  &$5.03(6)\times10^{-13}$  & - &0.99811(16) &0.2067(19) &$1.14(10)\times10^{-7}$
\\
\cite{abb+18a} & 0.92(3)  &$5.02(5)\times10^{-13}$  & $-4.0(13)\times10^{-16}$ & 0.99808(9) & 0.208(2) &$1.16(12)\times10^{-7}$
\\
\cite{pdd+19}  & 0.88(1)  &$5.05(3)\times10^{-13}$  & $-3.9(7)\times10^{-16}$ & 0.99807(6) & 0.209(1) &$1.04(6)\times10^{-7}$
\\
\cite{aab+20a} & 0.88(2)  &$5.09(3)\times10^{-13}$  & $-2.9(8)\times10^{-16}$ & 0.99794(7) & 0.210(2) &$1.10(9)\times10^{-7}$
\\
This work      & 0.861(13) &$5.1087(13)\times10^{-13}$ & $-2.61(55)\times10^{-16}$ & 0.998005(65) & 0.209(1) &$1.15(6)\times10^{-7}$ \\
\hline
    \end{tabular}
\end{table*}

The observed secular variation of orbital period in binary pulsars can include contributions from multiple effects \citep{dt91,lk05}. As for the case of PSR~J1909$-$3744 where the pulsar has negligible tidal interaction with the WD companion, the most relevant effects can be summarised below:
\begin{equation} \label{eq:Pbdot_terms}
\dot{P}^{\rm obs}_{\rm b}=\dot{P}_{\rm b}^{\rm Shk}+\dot{P}_{\rm b}^{\rm Gal}+\dot{P}_{\rm b}^{\rm GR}+\dot{P}^{\dot{m}}_{\rm b},
\end{equation}
The first term, $\dot{P}_{\rm b}^{\rm Shk}$, denotes the contribution from the transverse relative motion of the system to the solar system barycentre (SSB) which is known as the Shklovskii effect. Using the measured proper motion of the pulsar ($\mu_{\rm\alpha}$, $\mu_{\rm\delta}$) and its distance ($d$) derived from timing parallax, this term is estimated to be
\begin{equation} \label{eq:pbdot:Shk}
    \dot{P}^{\rm Shk}_{\rm b}=P_{\rm b}\left(\mu^2_{\rm\alpha}+\mu^2_{\rm\delta}\right)\frac{d}{c}=5.13(8)\times10^{-13},
\end{equation}
where $c$ is the speed of light. The second term, $\dot{P}_{\rm b}^{\rm Gal}$, is caused by the differential acceleration from the Galactic potential. As shown in \cite{dt91}, this effect can be calculated based on astrometric measurements and a Galactic potential model. Using our timing results and the model provided in \citet{McMillan:2017}, we have\footnote{The error in this calculation only reflects the error in the parallax. Accounting for a realistic error in the Galactic potential, one expects an error roughly an order of magnitude larger, which however, is still about an order of magnitude smaller than the error in $\dot{P}_{\rm b}^{\rm Shk}$.}
\begin{equation}
    \dot{P}_{\rm b}^{\rm Gal}=0.0265(6)\times10^{-13}.
\end{equation}
The third term, $\dot{P}_{\rm b}^{\rm GR}$, represents the contribution from gravitational wave damping in GR, and following \cite{pet64}, is found to be\footnote{Whenever being in combination with $T_\odot$, masses are understood in solar units.}
\begin{equation} \label{eq:pbdot:gw}
    \dot{P}_{\rm b}^{\rm GW}=-\frac{192\pi}{5}\left(\frac{2\pi}{P_{\rm b}}\right)^{5/3}\frac{(T_{\odot}m_{\rm c})^{5/3}q}{(q+1)^{1/3}}=-0.0279(3)\times10^{-13},
\end{equation}
where $T_{\odot} = (\mathcal{GM})_{\odot}^\mathrm{N} / c^3$, and $(\mathcal{GM})_{\odot}^\mathrm{N}$ is the solar mass parameter as defined by the IAU \citep{mamajek2015iau}. The last term, $\dot{P}^{\dot{m}}_{\rm b}$, is arising from the mass loss of the system. In the case of \psr, we consider the loss of rotational energy of the pulsar to be the dominant mass-loss effect, given that there is no expectation of significant stellar wind from a cooled WD with strong surface gravity like the companion of \psr~\citep[e.g.,][]{ung08}. Following \cite{dt91}, this can be worked out as\footnote{The uncertainty here only includes the measurement error of the parameters, except $I_{\rm p}$ whose value is assumed. It should be noted that Eq.~(\ref{eq:pbdot:mdot}) is an approximation for slowly rotating neutron stars, whose precision nevertheless is better than 10\% for PSR J1909$-$3744.}
\begin{equation} \label{eq:pbdot:mdot}
    \dot{P}^{\dot{m}}_{\rm b}=\frac{8\pi^2G}{T_{\odot}c^5}\frac{I_{\rm p}}{m_{\rm p}+m_{\rm c}}\frac{\dot{P}}{P^3}P_{\rm b}=4.85(6)\times10^{-16}.
\end{equation}
Here $I_{\rm p}$ and $\dot{P}$ denote the moment of inertia and the intrinsic period derivative of the pulsar, respectively. To obtain the estimate above, we used $I_{\rm p} = 1.4 \times 10^{45}$\,g\,cm$^2$, a typical value derived using our mass measurement and different equations of state that are in agreement with all pulsar and LIGO/Virgo constraints \citep{lp01,Capano2020}. We also recovered the intrinsic spin period derivative, by subtracting the contribution from the Shkloskii effect and the Galactic potential from the observed value as performed in the case of the orbital period derivative. These calculations are summarized in Table~\ref{tab:param} which shows that the intrinsic $\dot{P}$ of the pulsar is in fact more than a factor of five smaller than the observed value.

The analysis above clearly shows that the observed $\dot{P}_{\rm b}$ is dominated by the Shklovskii effect, next to which is the contribution from Galactic potential. Though the expected contribution from gravitational wave damping is well above the measurement uncertainty of the observed $\dot{P}_{\rm b}$, it is still below the estimation error of the Shklovskii contribution which mostly comes from the measurement uncertainty in parallax distance. Thus, with the current measurement precision $\dot{P}_{\rm b}^{\rm GW}$ is not separable from $\dot{P}^{\rm obs}_{\rm b}$ and we expect to be able to do so only when the precision in distance measurement is improved by at least a factor of 4 to 5. This could possibly be obtained by adding more existing data, extending the timing baseline, or using timing data from more sensitivity telescope \citep[e.g.,][]{bbb+18}. Subtracting all aforementioned contributions from $\dot{P}^{\rm obs}_{\rm b}$ gives $\dot{P}_{\rm b}^{\rm Exs} = (-1.7\pm7.8) \times 10^{-15}$ which is consistent with zero. This allows us to calculate the kinematic distance ($d_{\rm k}$) of the system with $\dot{P}^{\rm obs}_{\rm b}$ by directly assuming a zero excessive $\dot{P}_{\rm b}$ \citep{bb96}, which gives $d_{\rm k} = 1.157(3)$\,kpc, a highly consistent and more precise distance estimate.

The observed secular variation of projected semi-major axis in binary pulsars can also be induced by a series of effects as listed below \citep{lk05}:
\begin{equation} \label{eq:xdot}
\dot{x}^{\rm obs}=\dot{x}^{\rm Shk}+\dot{x}^{\rm Gal}+\dot{x}^{\rm GW}+\frac{d\epsilon_{\rm A}}{dt}+\dot{x}^{\rm SO}+\dot{x}^{\rm PM}.
\end{equation}
The first term, $\dot{x}^{\rm Shk}$, stands for the contribution from the Shklovskii effect, and using our timing astrometric measurements, is estimated to be:
\begin{equation} \label{eq:xdot:Shk}
    \dot{x}^{\rm Shk}=x\left(\mu^2_{\rm\alpha}+\mu^2_{\rm\delta}\right)\frac{d}{c}\simeq7.3\times10^{-18}.
\end{equation}
The second term, $\dot{x}^{\rm Gal}$, is the contribution from the differential Galactic potential, and following the same approach as for $\dot{P}_{\rm b}^{\rm Gal}$, is found to be
\begin{equation} \label{eq:xdot:Gal}
\dot{x}^{\rm Gal}\simeq 3.8\times10^{-20}.
\end{equation}
The third term, $\dot{x}^{\rm GW}$, is caused by gravitational wave damping. Following \citep{pet64} and using our mass measurements, it is estimated to be
\begin{equation}
    \dot{x}^{\rm GW}=-x\frac{64}{5}\left(\frac{2\pi}{P_{\rm b}}\right)^{8/3}\frac{(T_{\odot}m_{\rm c})^{5/3}q}{(q+1)^{1/3}} \simeq -2.8\times10^{-20}.
\end{equation}
The fourth and fifth term, $d\epsilon_{\rm A}/dt$\footnote{Here $\epsilon$ is defined as in Eq.~(2.6) of \cite{dt92}.} and $\dot{x}^{\rm SO}$, denote contribution from geodetic precession of the pulsar spin axis and the precession of the orbital plane due to spin-orbit coupling effect (classical and/or Lense-Thirring), respectively. Following \cite{dt91}, $d\epsilon_{\rm A}/dt$ is of order $\Omega^{\rm geo}\cdot P/P_{\rm b}\sim10^{-19}$. The spin-orbit precession induced by the companion's rotation has so far been measured in a small number of different binary pulsar systems \citep{kbm+96,sjm14,vbv+20}. However, spin-orbit contributions to $\dot{x}$ vanish when the spin axes of the pulsar and companion are aligned with the orbital angular momentum \citep[e.g.,][]{dt92,wex98}, which is expected to be the case for fully recycled binary pulsars such as \psr{} from the binary evolution aspect \citep[e.g.,][]{tau12}. Additionally, the pulsar is in a relatively wide orbit with the WD companion, which will further suppress the spin-orbit coupling effects by a few orders of magnitude compared with the case of PSR~J1141$-$6545 studied in \citep{vbv+20}.

\begin{figure}
  \centering
  \hspace*{-0.4cm}
  \includegraphics[scale=0.6]{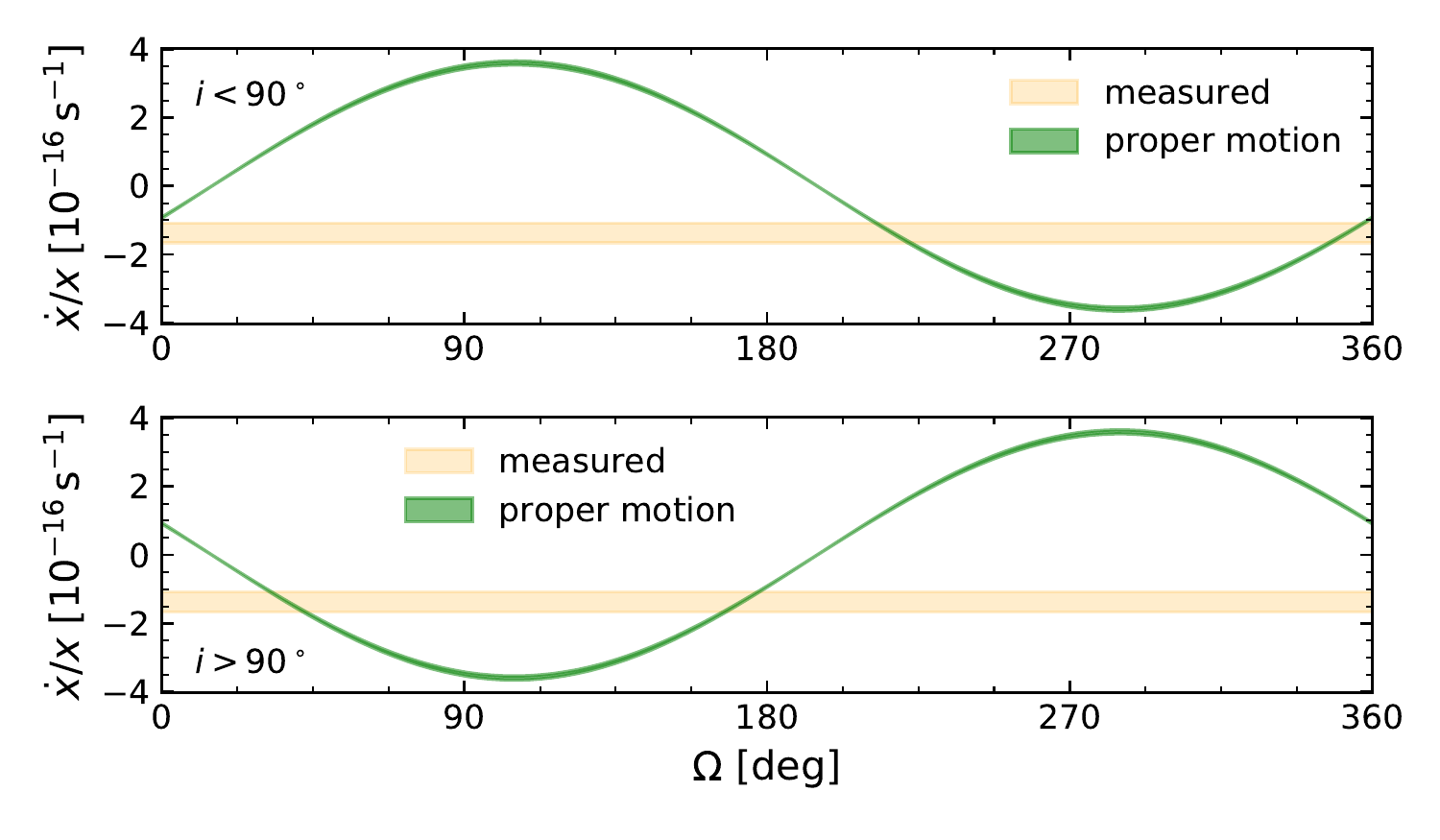}
  \caption{Contribution of the proper motion to $\dot x/x$ as a function of $\Omega$, for $i<90^\circ$ (upper) and $i>90^\circ$ (lower) using the measured $\sin{i}$. The observed value of $\dot x/x$ and its 1-$\sigma$ uncertainty are shown with the horizontal orange band. The four intersections represent the derived $\Omega$ values.
\label{fig:xdot}}
\end{figure}

Therefore, it can be concluded that the contribution from the relative motion between the pulsar and the SSB, $\dot{x}^{\rm PM}$, should be dominating in $\dot{x}^{\rm obs}$. Following the derivations presented in \citet{ajrt96} and \citet{kop96}, this effect can be expressed as\footnote{Note that there is a conversion between the $\Omega$ and $i$ here and those in \cite{kop96}: $\Omega = \pi/2 - \Omega_{\rm K96}$, $i=\pi-i_{\rm K96}$.}
\begin{equation} \label{eq:xdot:PM}
  \dot{x}^{\rm PM}=x\left(\mu_{\alpha} \cos \Omega-\mu_{\delta} \sin \Omega\right) \cot i \,.
\end{equation}
It can be seen that one can derive the ascending node of the binary system, $\Omega$, once $\dot{x}^{\rm PM}$ and $i$ are measured. This is shown in Fig.~\ref{fig:xdot}, where we plot the contribution to $\dot x/x$ from the proper motion for PSR~J1909$-$3744 as a function of $\Omega$, in conjunction with the measured $\dot{x}/x$. Accordingly, we have identified four possible solutions considering the ambiguity in $i$ ($\sin{i}$ equals to the shape of Shapiro delay, $s$): $\Omega=217\pm5$, $352\pm5$\,deg when $i<90$\,deg, and $\Omega=37\pm5$, $172\pm5$\,deg when $i>90$\,deg. We noted that \cite{rhc+16} reported only one possible solution of such and \cite{aab+20a} reported two. In principle, this ambiguity can be resolved if the annual-orbital parallax of the binary system is measurable, via directly fitting for the ``Kopeikin'' (KOM and KIN) parameters in the \textsc{tempo2} software package as e.g., has been carried out in the PSR~J1713+0747 system \citep[see e.g.][]{dcl+16}. However, with the formula in \cite{kop95}, we estimate the scale of the feature in the timing data induced by annual-orbital parallax to be $\sim 6$\,ns, well below any of the published timing precision of PSR~J1909$-$3744. This is mostly due to the fact that PSR~J1909$-$3744 has a small and edge-on orbit with a distance on order of kpc. In practice, the Kopeikin parameters can also be highly correlated with some of the Keplerian parameters (e.g., the orbital period), which would further suppress the subtractable feature from the fit. Thus, our current timing precision does not allow to distinguish among the four possible solutions by modelling the annual-orbital parallax of the system. This has been verified by our attempt to switch on the fit for the Kopeikin parameters in our analysis, where all of the four solutions can be found, resulting in the same post-fit residual rms and returning the same goodness-of-fit. We however note that studying the orbital variations in the interstellar scintillation of the pulsar could potentially provide additional constraint over orbital inclination and help to resolve the ambiguity in our solutions \citep[e.g.,][]{lyn84,rch+19}.

\subsection{Binary evolution}
\label{ssec:binary}
\psr\; is an archetypical binary MSP with an orbital period of $P_{\rm b}=1.53\;{\rm days}$, a helium-core WD companion of mass $m_{\rm c}\simeq0.21 $\,$M_\odot$, a (very) low eccentricity $e=1.15\times 10^{-7}$, a rapid NS spin of $P=2.9\;{\rm ms}$, a small spin period derivative of $\dot{P}=1.4\times 10^{-20}$ (yielding a surface magnetic dipole field of approximately $B\simeq 1-2\times 10^{8}\;{\rm G}$), and a NS mass of $m_{\rm p}\simeq1.49$\,$M_\odot$. All these characteristics imply a plain vanilla formation scenario from a progenitor system which was a low-mass X-ray binary (LMXB) with a NS accretor and a main-sequence donor of mass $1.0-2.3\;M_\odot$ \citep{tv06,tau11}. Donor stars less massive than about $1.0\;M_\odot$ (depending on metallicity) will not terminate their evolution within a Hubble time, and stars more massive than $\sim 2.3\;M_\odot$ start to burn helium without their core becoming degenerate.

\begin{figure}
\centering
\includegraphics[scale=0.5]{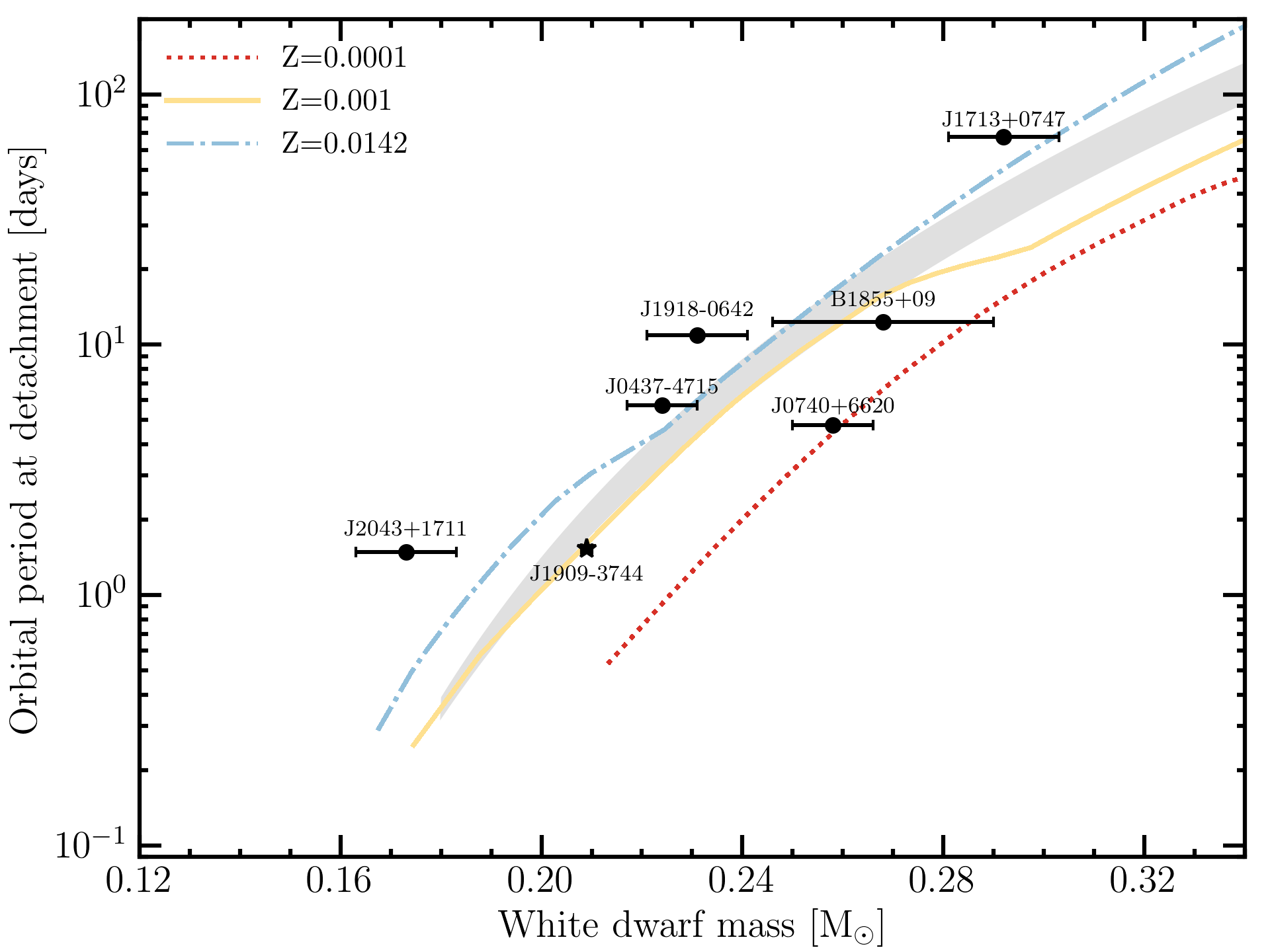}
\caption{Orbital period at the end of the LMXB phase versus the mass of the proto-WD. The lines represent the mass-period relation resulting from our detailed evolutionary models for Z=0.0001 (dotted red), Z=0.001 (solid yellow) and Z=0.0142 (dashed-dotted blue), respectively. The grey region represents the mass-period relation as calculated by \citet{ts99a}. The circles stand for various He WD companions of MSPs for which the mass  was determined from the Shapiro delay. The star represents the observed data for \psr{}.}.
\label{fig:mass_period}
\end{figure}

In order to study the formation and evolution of \psr, we employed new binary evolutionary models computed with the open source stellar evolution code MESA \citep{paxton2011, paxton2013, paxton2015, paxton2018}, version 12115, with the assumed physics similar as presented in \cite{imt+16} (e.g including rotational mixing and element diffusion). The new evolutionary models will be soon available publicly (Istrate et. al 2020, in prep). For the initial binary conditions, we assumed a 1.1~$M_{\odot}$ donor mass, 1.3~$M_{\odot}$ NS accretor treated as a point mass, and an accretion efficiency of $\epsilon \sim 0.23$ ($\beta=0.77$).

From a stellar evolution point of view, one expects a tight correlation between the orbital period at the end of the LMXB phase and the mass of the proto-He WD  \citep[e.g.][]{sav87,joss1987,rappaport1995,ts99a,lin2011,jia2014,itla14}. This results from the relatively tight correlation existing between the degenerate-core mass of a red giant star and its radius \citep{refsdal1971}.
The mass-period relation depends primarily on the assumed metallicity and to some extent on other stellar parameters, such as the adopted mixing length value, $\alpha_{\mathrm{MLT}}$ and partly on the initial donor mass \citep{ts99a}.
Our models presented below assume $\alpha_{\mathrm{MLT}}=2.0$.

\begin{figure}
\centering
\includegraphics[scale=0.5]{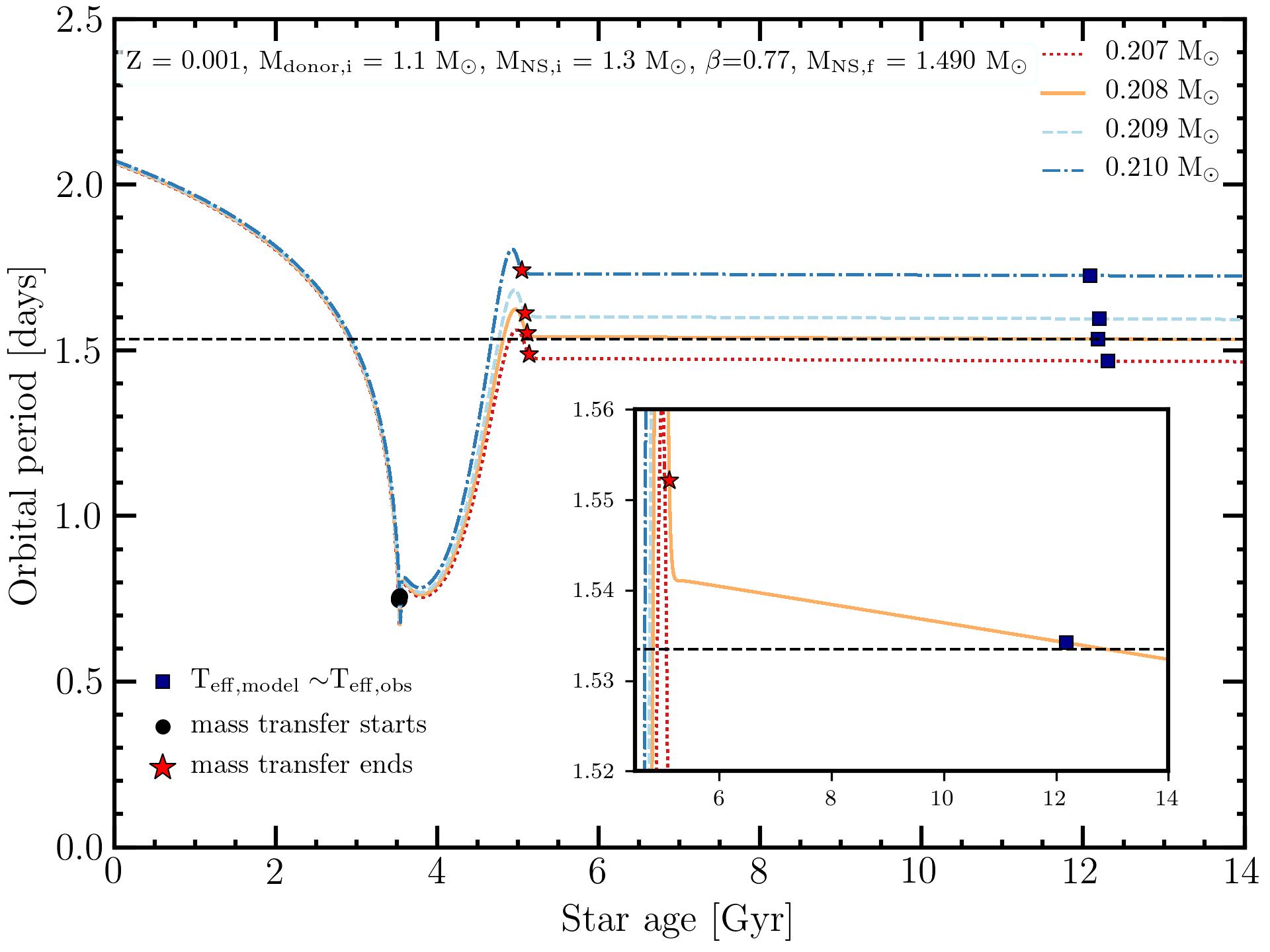}
\caption{Orbital period evolution for selected tracks resulting in WD masses of 0.207, 0.208, 0.209 and 0.210~$M_{\odot}$, respectively. The initial binary configuration consists of a 1.1~$M_{\odot}$ donor mass and a 1.3~$M_{\odot}$ NS with an accretion efficiency of $\epsilon$ $\sim 0.23$ at Z=0.001. The horizontal dashed line marks the observed orbital period of \psr{}. With the assumed $\epsilon$, the final NS mass is $\sim 1.49$~$M_{\odot}$, in agreement with the measured value.}
\label{fig:orbital_evolution}
\end{figure}

Fig.~\ref{fig:mass_period} shows the orbital period at the end of the LMXB phase versus the mass of the He~WD for Z=0.0001, Z=0.001 and solar-like metallicity of Z=0.0142.
The gray area represents the fitted mass-period relation for population I and II stars from \cite{ts99a}. Over-plotted are several observed values for MSP companions for which the mass\footnote{The values are taken from  \url{https://www3.mpifr-bonn.mpg.de/staff/pfreire/NS_masses.html}.} of the He~WD was measured from the Shapiro delay and therefore it is independent of the WD cooling models used. One should note that the observed mass-period relation can appear to be different from the theoretical mass-period relations in Fig.~\ref{fig:mass_period}, especially at relatively small orbital periods or alternatively small WD masses. The most important contributor to this discrepancy is the shrinkage of the orbit due to gravitational wave radiation. There are also smaller effects due to the change in the orbit as well as in mass of the WD caused by the occurrence of hydrogen shell flashes, which are stronger at higher metallicities \citep[see for example the discussion in][]{mik+20}. Finally, deviations can occur due to pulsar irradiation of the donor star which results in a widening of the orbit while the WD mass decreases \citep{krst88,srp92,vnvj12,ccth13}.

The observed orbital period and companion mass for \psr\; are marked with a black star and are in agreement with the values predicted for Z=0.001. Notably, the companion of \psr\; has one of the most accurate mass measurements from the Shapiro delay. As previously mentioned, the theoretical mass-period relations can vary based on assumed value of mixing length, $\alpha_{\mathrm{MLT}}$.  
For example, it would be possible that one could find a solution for a somewhat lower metallicity than Z=0.001 with $\alpha_{\mathrm{MLT}}$ smaller than 2.0 or slightly higher metallicity than Z=0.001 with $\alpha_{\mathrm{MLT}}$ larger than 2.0. The value of $\alpha_{\mathrm{MLT}}$ is relatively uncertain and possibly is degenerate with stellar mass or metallicity \citep[e.g][]{tayar2017,viani2018,sonoi2019, valle2019}.

Fig.~\ref{fig:orbital_evolution} shows the orbital period evolution for a few selected evolutionary tracks with Z=0.001 that produce a final WD mass of 0.207, 0.208, 0.209 and 0.210~$M_{\odot}$, respectively. The measured companion mass of \psr\; is consistent with a 0.208, 0.209 and a 0.210~$M_{\odot}$ solution; however, as it can be seen from Fig.~\ref{fig:orbital_evolution}, the measured orbital period is only consistent with the 0.208\,$M_{\odot}$ track. Notably, at the moment of the Roche-lobe detachment, the orbital period was slightly larger, $\sim 1.55\;{\rm days}$, but after $\sim 7\;{\rm Gyr}$ of orbital decay due to gravitational wave radiation it crosses the present value of 1.53~days (see inlet in Fig.~\ref{fig:orbital_evolution}).

In order to obtain the final NS mass in agreement with the observed value ($m_{\mathrm{p}}\sim 1.49 M_{\odot}$), we fine-tuned the accretion efficiency, $\epsilon$ to a value of $\sim 0.23$, given our assumed NS birth mass of $1.3\;M_\odot$ and initial donor mass of $1.1\;M_\odot$. This value is close to the upper limit of the accretion efficiency from the following reasons: a higher initial donor mass implies a smaller value of $\epsilon$, and a smaller initial donor mass (which would imply a larger value for $\epsilon$) is unlikely as the evolution of the system to the current observed values will take longer than the Hubble time.

\begin{figure}
\centering
\includegraphics[scale=0.5]{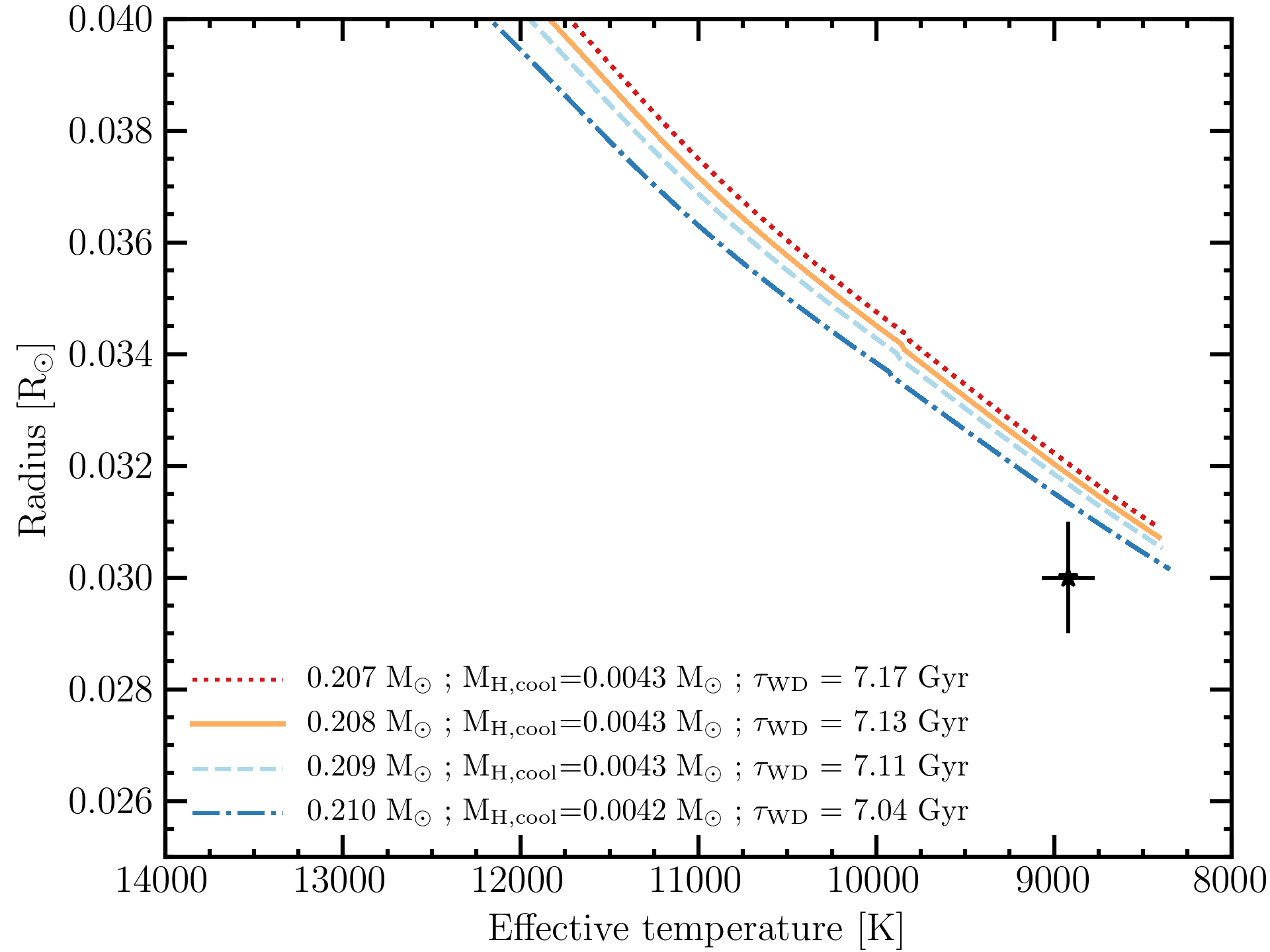}
\caption{Radius versus effective temperature for WD cooling tracks corresponding to the evolutionary tracks (Z=0.001) shown in Fig.~\ref{fig:orbital_evolution}. The black star represents the observed values. The labels denote the WD mass, the total amount of hydrogen available at the beginning of the cooling track (defined as the maximum effective temperature) and the associated WD age (from the end of the LMXB phase until it reaches the observed value of $T_{\mathrm{eff}}=8920$\;K).}
\label{fig:logg_teff}
\end{figure}

\begin{figure}
\centering
\includegraphics[scale=0.5]{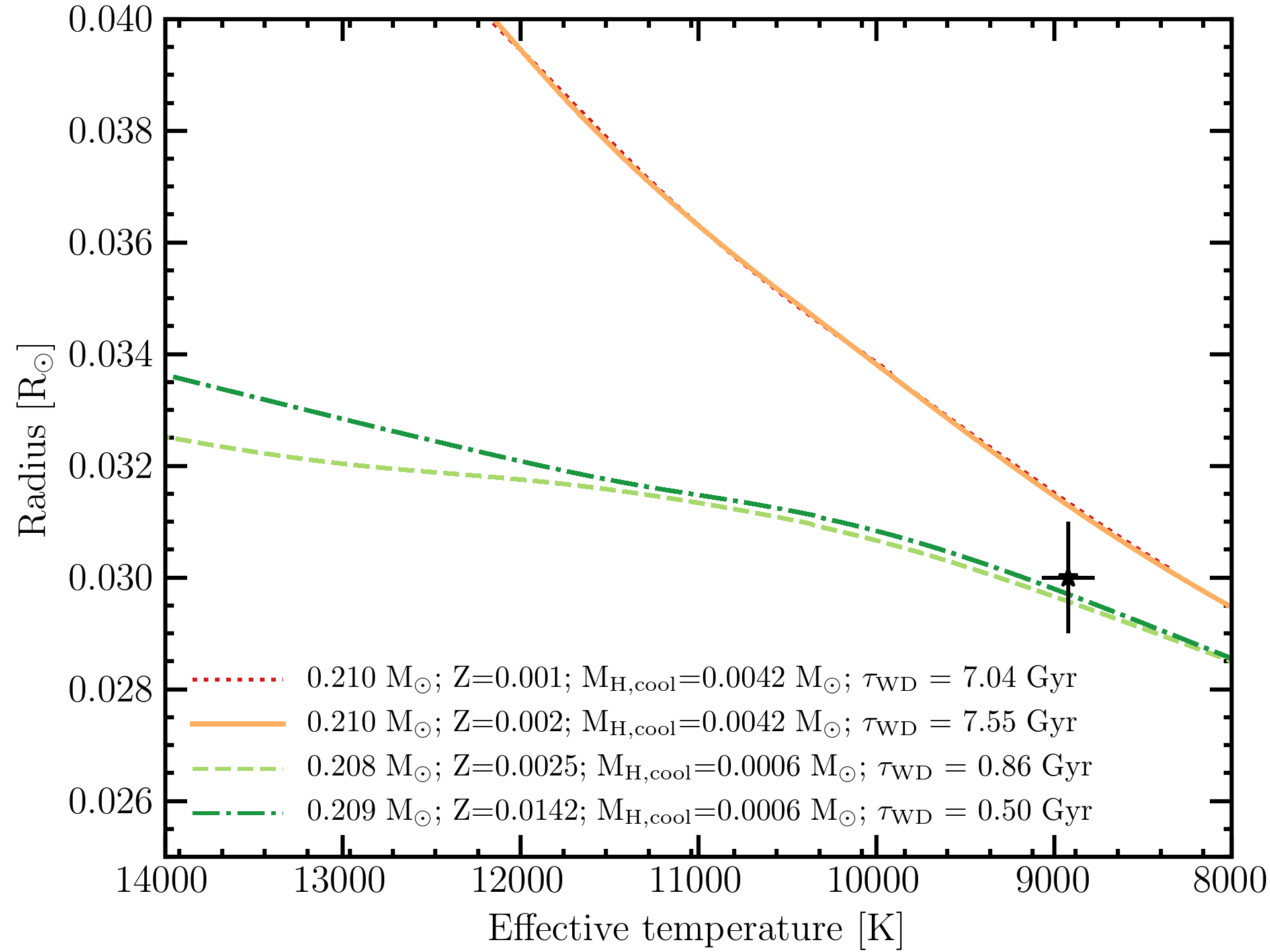}
\caption{Radius versus effective temperature for WD cooling tracks resulting from various metallicities with comparable mass to that of the companion of \psr{}, $m_{\mathrm{c}}\sim 0.21M_{\odot}$. The two families of tracks represent systems that do not experience hydrogen flashes (upper tracks with a thick envelope, Z=0.001 and Z=0.002) and systems that undergo flashes (bottom tracks with a thin envelope, Z=0.0025 and 0.0142 respectively).}
\label{fig:logg_teff_metallicity}
\end{figure}

It has been demonstrated that characteristic ages, $\tau =P/2\dot{P}$ are often very poor true age estimators for recycled pulsars \citep{tau12,tlk12}. Therefore, we use the
{\color{black}cooling} age of the WD, $\tau_{\mathrm{WD}}$ (defined as the time elapsed since the end of the LMXB phase until the WD reaches the observed effective temperature) to
estimate the age of the \psr\;system. The WD companion has a measured radius of 0.030$\pm$0.001\;R$_{\odot}$\citep{ant13} and an effective
temperature of 8920$\pm$150\;K \citep{kilic2018}. Fig.~\ref{fig:logg_teff} shows the evolution of the radius versus effective temperature (cooling evolution) for the selected
evolutionary models presented above. None of the cooling tracks in Fig.~\ref{fig:logg_teff} are consistent at the 1-$\sigma$ level with the observed value of the WD radius. These cooling tracks are all characterized by a thick hydrogen envelope of $\sim 4\times 10^{-3}$~$M_\odot$ with an associated WD age of the order of 7~Gyr. We therefore search for another solution to see if we can better match the derived WD radius.

The most important parameter that determines the cooling history of a WD, for a given mass, is the mass of its envelope. Helium-core WDs can be divided into thin or thick hydrogen
envelopes depending on whether or not they experience diffusion-induced hydrogen shell flashes \citep[e.g][]{althaus2001,imt+16} with a large impact on their cooling ages \citep[the {\em cooling dichotomy}, e.g.][]{asvp96,vbjj05,itla14}. The mass threshold for a hydrogen flash occurrence depends strongly on the assumed metallicity \citep{imt+16}. This is also demonstrated in Fig.~\ref{fig:logg_teff_metallicity}, where we show our computed cooling tracks for a helium-core WD of $\sim 0.21$~$M_\odot$ resulting from initial donors with a metallicity of Z=0.001, Z=0.002, Z=0.0025 and Z=0.0142, respectively. The mass threshold for flashes to occur at the metallicity suggested by the binary evolution, Z=0.001, is $\sim 0.23$~$M_\odot$. This threshold drops to $\sim 0.215$~$M_\odot$ for Z=0.002. As a consequence, the upper two tracks with Z=0.001 and Z=0.002 have thick envelopes ($M_{\mathrm{H, cool}}\sim 4 \times10^{-3}$~$M_\odot$, and no flashes) and the bottom tracks have thin hydrogen envelopes ($M_{\mathrm{H, cool}}\sim \textbf{6}\times 10^{-4}$~$M_\odot$, and undergo flashes). The observed properties of the \psr\; companion are consistent at the 1-$\sigma$ level only with thin hydrogen envelope models with a corresponding age of $\sim 0.5\;{\rm Gyr}$. However, at the 2-$\sigma$ level, both thick and thin hydrogen envelope solutions are possible, with a resulting large spread in the cooling age from 7~Gyr to 0.5~Gyr, respectively. Given the current uncertainties of the radius, one cannot firmly derive the cooling age of the WD and therefore the age of the pulsar.

We stress here that the binary evolutionary models leading to a He WD orbiting a MSP use various parameters that are not well constrained yet.
At the level of precision that we investigated in this work, parameters such as $\alpha_{\mathrm{MLT}}$, overshooting, initial donor mass,  efficiencies of various mixing processes, just to name a few, play an important role in determining the precise mass-period relation at a given metallicity. Varying these parameters, one could obtain a thin hydrogen envelope solution self-consistently from the binary evolution in the same manner as the solution presented for the thick envelopes.
Alternatively, the mass threshold for flash occurrence  as well as the amount of hydrogen remaining after the hydrogen-shell flash episodes might also depend on other physical parameters besides the well-known dependence on  element diffusion \citep[e.g ][]{althaus2001,itla14, imt+16}). Once a more precise  and reliable radius measurement can be obtained, the \psr\; system will serve as a powerful benchmark for investigating the influence and calibrating parameters that otherwise are thought to have only second-order effects both for the binary evolution and  cooling of helium-core WDs.

To summarise, we have obtained a binary solution in agreement with the observed orbital period, WD mass and NS mass for Z=0.001, with an upper limit for the accretion efficiency $\epsilon \sim 0.23$. However, this solution is not self-consistent with the cooling properties of the \psr\; WD companion. The radius, effective temperature and mass of the WD can best be explained by models with a thin hydrogen envelope (most likely with a chemical abundance Z $> 0.001$) with a corresponding age of only $\sim 0.5\;{\rm Gyr}$, but at the 2-$\sigma$ level, the cooling age varies from 7~Gyr to 0.5~Gyr. 
We emphasise here that the WD companion to \psr\; is a perfect showcase for the cooling dichotomy of He WDs, which makes an age determination very difficult in practice. To pinpoint with certainty a cooling age, one would need a more accurate radius measurement. Currently, this is mostly limited by uncertainties in the reddening estimate, and the precision of available photometric data. Therefore, a significant improvement would require deeper imaging observations at multiple wavelengths and a more precise extinction model.

Fortunately, the effective temperature and the surface gravity of the WD companion of  \psr\;  place it closely to the red edge of the instability strip of extremely-low mass  white dwarf variable stars (ELMVs) \citep{corsico2012,grootel2013,kilic2018}. Additional constraints on the thickness of the hydrogen envelope  (and therefore cooling age) could be obtained if one would investigate the asteroseismological properties for the possible evolutionary solutions and (ideally) compare it with observed pulsation periods \citep{calcaferro2018}. This is beyond the scope of this paper and it will be addressed in a future work.

\subsection{3D velocity and Galactic motion}
\label{ssec:motion}
\vspace{2mm}

PSR~J1909$-$3744 is one of the few pulsars where we have full information on the 3D spatial velocity with respect to the SSB. The proper motions in right ascension and declination are known with high precision from timing observations (see Tab.~\ref{tab:param}). The same is the case for the distance to the PSR~J1909$-$3744 system. In particular, the kinematic distance $d_{\rm k}$ has an uncertainty of less than 0.3\%. Combining this information gives a transverse velocity of $203.2 \pm 0.5\,{\rm km\,s^{-1}}$. Finally, from high-resolution spectroscopy \cite{ant13} was able to infer a systemic radial velocity of $-73\pm30\,{\rm km\,s^{-1}}$ with respect to the SSB. The corresponding total velocity with respect to the SSB is found to be $218 \pm 10\,{\rm km\,s^{-1}}$.

With the local (SSB frame) position and 3D velocity of PSR~J1909$-$3744 at hand, we can now reconstruct the Galactic motion of the PSR~J1909$-$3744 system. For this purpose, we make use of the Galactic gravitational potential given by \cite{McMillan:2017}. We introduce a Galactic frame centered at the location of the Galactic center, with the $X-Y$ plane coinciding with the Galactic plane, and the Sun located at $X = -8.2$\,kpc and $Y =  Z = 0$\,kpc. For the purpose of this section, we can safely ignore the small and somewhat uncertain offset of the Sun from the Galactic plane \citep[see e.g.][]{ymw17}. The $X$ value for the Sun corresponds to the Galactic center distance $R_0$ used in the \cite{McMillan:2017} model, which is in good agreement with the latest distance estimate for Sgr A$^\ast$ by the GRAVITY Collaboration \citep{GRAVITY:2020}. To calculate today's velocity of the PSR~J1909$-$3744 system with respect to our $X-Y-Z$ frame, we need the Galactic velocity vector of the SSB. For a given $R_0$, the proper motion measurements for Sgr A$^\ast$ by \cite{Reid:2004} can directly be converted into the $Y$- and $Z$-velocity of the SSB. For $R_0 = 8.2$\, kpc we find $V_Y^{\rm SSB} = 248\,{\rm km\,s^{-1}}$ and  $V_Z^{\rm SSB} = 8\,{\rm km\,s^{-1}}$. Furthermore, we adopt $V_X^{\rm SSB} = 9\,{\rm km\,s^{-1}}$, in agreement with \cite{McMillan:2017} as well as \cite{Reid:2014}. As a result, we find the following Galactic velocity for PSR~J1909$-$3744
\begin{eqnarray}
  V_X &=& -67 \pm 28 \,{\rm km\,s^{-1}} \,,\label{eq:VX}\\
  V_Y &=& 46 \pm 1  \,{\rm km\,s^{-1}} \,, \label{eq:VY}\\
  V_Z &=& 16 \pm 10 \,{\rm km\,s^{-1}} \,. \label{eq:VZ}
\end{eqnarray}
The small error in $V_Y$ does not come as a surprise. On one hand, the error budget in the velocity is clearly dominated by the error in the observed radial velocity; on the other, PSR~J1909$-$3744 is located at $l = 359.7\deg$ and $b = -19.6\deg$ (see Tab.~\ref{tab:param}), meaning that the radial velocity has practically no component in $Y$-direction. Furthermore, it is interesting to note how small $V_Y$ is, compared to the motion of the SSB, i.e.\ $V_Y^{\rm SSB} = 248\,{\rm km\,s^{-1}}$, although PSR~J1909$-$3744 is practically in our Galactic neighbourhood. From this, one can already assess that the PSR~J1909$-$3744 must have a somewhat peculiar Galactic motion. The velocity with respect to its local standard of rest is therefore comparatively large, amounting to $201 \pm 10\,{\rm km\,s^{-1}}$.

\begin{figure}
\centering
\includegraphics[width=\columnwidth]{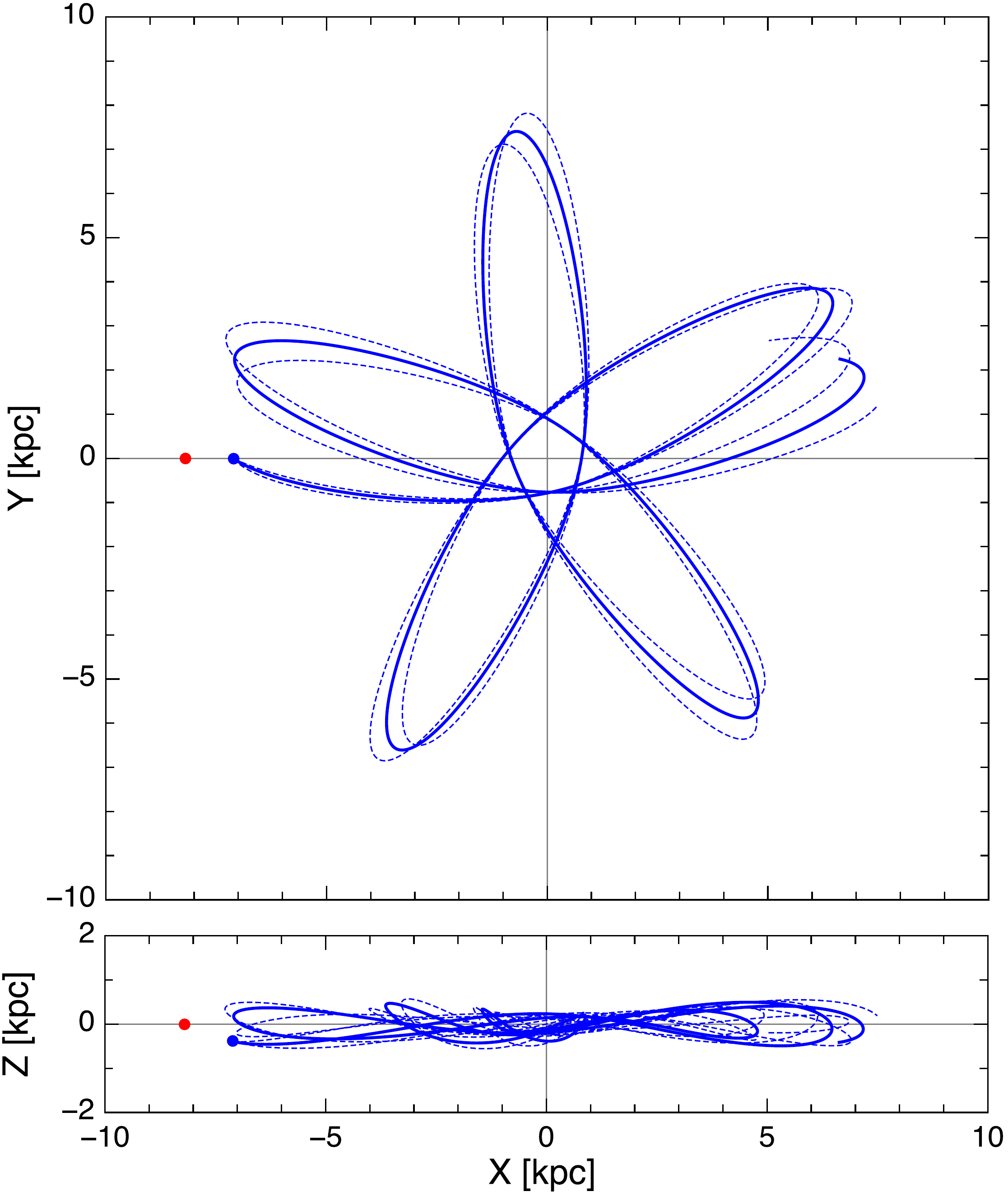}
\caption{Galactic motion of the PSR J1909$-$3744 system, starting 500\,Myr in the past. Today's location is indicated by a blue circle. The thin dashed lines correspond to a variation of the radial velocity $V_r$ by plus/minus one standard error. The potential for the Galactic gravitational field has been taken from \protect\cite{McMillan:2017}. The location of the Sun is indicated by a red circle.
\label{fig:GalMotion}}
\end{figure}

We have used the software package provided by \cite{McMillan:2017} to integrate the Galactic motion of the PSR~J1909$-$3744 system back in time, using its current location and inverting its current velocity in Eqs.~(\ref{eq:VX})--(\ref{eq:VZ}). The result is given in Fig.~\ref{fig:GalMotion}, where we have stopped the integration at 500\,Myr, which is below the expected age of the pulsar, i.e.\ long after the supernovae that formed the neutron star and imparted a kick to the system (see Section~\ref{ssec:binary}). As one can see, PSR~J1909$-$3744 is on a highly eccentric orbit, currently being close to its maximum distance from the Galactic center. About 30\,Myr ago it had its closest approach to the Galactic center with a distance of $\lesssim 1$\,kpc. The motion is confined to a region comparably close to the Galactic plane with $|Z| \lesssim 0.5$\,kpc. We have used different models for the Galactic potential (provided with the software package of \cite{McMillan:2017}) to verify the robustness of our finding against uncertainties in the Galactic gravitational potential. The fine details of the Galactic orbit of PSR~J1909$-$3744 should generally be taken with a grain of salt, since already the assumption of an axisymmetric mass distribution in our Galaxy is only a first order approximation. For instance, in the inner few kpc one has the central bar as a clearly non-axisymmetric structure \citep[see][and references therein]{Bland-Hawthorn:2016}. Although this will modify the details of the orbit, it is not expected to change the overall picture that the Galactic orbit of PSR~J1909$-$3744 is highly eccentric and takes PSR~J1909$-$3744 into the inner Galactic region.

\subsubsection{On the small eccentricity of \psr}
\label{sssec:ecc}
The orbital eccentricity of \psr\; is a record low among all known binary pulsars with a value of just $e=1.15\times 10^{-7}$.
Given a potential age of from 0.5 to 7\,Gyr of this binary pulsar (inferred from the WD cooling age, as discussed in Section~\ref{ssec:binary}), it is of interest to investigate whether we can use it as a probe of the stellar density in the Galactic disk. It is anticipated that orbital eccentricities might be induced by dynamical interactions with field stars, similar to the examples of resulting high eccentricities seen among some binary pulsars in dense stellar environments like globular clusters. The question is if for an assumed old system like \psr\; (depending on the exact WD cooling models discussed in Section~\ref{ssec:binary}), the column density of field stars within its cross-section radius, accumulated from its motion in the disk (Section~\ref{ssec:motion}) over its lifetime of several Gyr, may reach a critical level.

\citet{rh95} investigated resulting eccentricities of binary MSPs induced via such dynamical interactions, and they derived the following expression (their Eq. 6):
\begin{equation}
e\simeq \left(\frac{\eta}{400}\right)^{5/2}\,P_{\rm b}^{5/3}
\end{equation}
where $\eta \equiv t_9\,n_4/v_{10}$, and $t_9$ is the age in units of Gyr, $n_4$ is the stellar density (all assumed to be of mass $1\;M_\odot$) in units of $10^4\;{\rm pc}^{-3}$, and $v_{10}$ being the one-dimensional velocity dispersion in units of $10\;{\rm km\,s}^{-1}$.
Given $P_{\rm b}=1.53\;{\rm days}$ and $e=1.15\times 10^{-7}$, it yields $\eta\simeq 0.50$. Assuming $v_{10}\simeq 5$ (based on the relative velocity between \psr\; and local field stars), and $t_9\simeq 10$ as an extreme upper limit, we find that $n_4\la 0.25$. I.e. the critical number density needed to explain the small observed eccentricity of $1.15\times 10^{-7}$ for \psr\; based on dynamical interactions is about $2\,500\;{\rm stars\;pc}^{-3}$. This value is four orders of magnitude larger that the number density of stars in the solar neighborhood, $n_\odot = 0.12\;{\rm pc}^{-3}$ \citep[e.g.,][]{hf00}.

Therefore, we conclude that it is not surprising that \psr\; was able to retain its small eccentricity although potentially plowing through the Galactic disk for possibly up to several Gyr. We also note that this eccentricity is expected to be a residual value from its formation process via mass transfer from the progenitor star of the WD \citep{phi92}.


\subsection{Tests of alternative gravity theories}
\label{ssec:test}

\subsubsection{Testing for dipolar radiation}

Short orbital period ($P_{\rm b} \lesssim 1$\,d) pulsar-WD systems have turned out to provide some of the most stringent limits on dipolar gravitational radiation, a prediction by many alternative gravity theories \citep{wex14}. With its orbital period of approximately 1.5 days, \psr\ is rather slow compared to the systems like PSR~J1738+0333 \citep{fwe+12} or PSR~J0348+0432 \citep{afw+13}. However, its outstanding timing precision and the corresponding precise measurement of $\dot{P}_{\rm b}$ makes it still interesting for a test of dipolar gravitational wave damping. Moreover, while for PSR~J1738+0333 and PSR~J0348+0432 optical measurements and the modelling of WD spectra were required to obtain the masses of pulsar and companion, for \psr\ we can estimate these masses directly from the high precision timing observations due to the presence of a prominent Shapiro delay.

Here we discuss our dipolar radiation test within the framework of Bergmann-Wagoner theories. Bergmann-Wagoner theories represent the most general mono-scalar–tensor theories that are at most quadratic in the derivatives of the fields in their action \citep{will18}. Quite a number of well known mono-scalar-tensor theories belong to this class, like Jordan-Fierz-Brans-Dicke (JFBD) gravity \citep{Jordan:1955schwerkraft, Fierz:1956,Brans:1961}, DEF gravity \citep{de93,de96}, MO gravity \citep{Mendes:2016}, $f(R)$ gravity \citep{De_Felice:2010}, and massive Brans-Dicke gravity \citep{Alsing:2012}.

In Bergmann-Wagoner theories, the field equations for the (physical) ``Jordan-frame metric'' $g_{\mu\nu}$ and the scalar field $\phi$ can be derived from the following action
\begin{eqnarray}
  S &=& \frac{1}{16\pi G_0}\int\sqrt{-g}\,d^4x \left(\phi R -
  \frac{\omega(\phi)}{\phi} \partial_\mu\phi\partial^\mu\phi - U(\phi)\right)
  \nonumber\\ &&
  + S_{\rm mat}\left[\psi_{\rm mat}^A,g_{\mu\nu}\right] \,,
\end{eqnarray}
where $G_0$ denotes the fundamental (``bare'') gravitational constant, $g$ is the determinant of the metric $g_{\mu\nu}$, and $R$ the curvature scalar. The two functions $\omega(\phi)$ and $U(\phi)$ denote the coupling function and the scalar potential, respectively. Here we will assume that the influence of $U(\phi)$ is negligible on the relevant scales \cite[cf.][]{Alsing:2012}. $S_{\rm mat}$ is the action of the matter fields $\psi_{\rm mat}^A$, which couple universally to the spacetime metric $g_{\mu\nu}$.

The Newtonian gravitational constant, as measured in a Cavendish-type
experiment, is given by
\begin{equation}
    G_{\rm N} = \frac{G_0}{\phi_0(1 - \zeta)} \,,
\end{equation}
where $\phi_0$ is the cosmological background field and $\zeta \equiv 1/(2\omega(\phi_0) + 4)$. A further quantity, which we will need below, is the sensitivity
\begin{equation}
    s_i \equiv \left(\frac{d\ln m_i(\phi)}{d\ln\phi}\right)_{\phi_0} \,,
\end{equation}
which accounts for the dependence of the mass of the $i$-th body (for a fixed number of baryons) on a change in its ambient scalar field \citep{will18}.

For our dipolar radiation test with \psr\ we can utilize three post-Keplerian parameters. These are the {\em range}
\begin{equation}\label{eq:rShap}
   r = (1 - \zeta) \, \tilde m_{\rm c} \simeq
       \tilde m_{\rm c}\,,
\end{equation}
and {\em shape}
\begin{equation}\label{eq:sShap}
   s = x\left(\frac{2\pi}{P_{\rm b}}\right)^{2/3}
        \frac{(\tilde m_{\rm p} + \tilde m_{\rm c})^{2/3}}
             {(1-2\zeta s_{\rm p})^{1/3}\tilde m_{\rm c}} \,,
\end{equation}
of the Shapiro delay caused by the WD companion, and a potential (intrinsic) change of the orbital period due to scalar dipolar radiation
\begin{equation} \label{eq:Pbdot}
    \dot{P}_{\rm b} \simeq
    \dot{P}_{\rm b}^{\rm dipole} \simeq
    -\frac{16\pi^2}{P_{\rm b}} \,
     \frac{\tilde m_{\rm p} \tilde m_{\rm c} }
          {\tilde m_{\rm p} + \tilde m_{\rm c}} \,
     \zeta s_{\rm p}^2 \,,
\end{equation}
where $\tilde m_i \equiv G_{\rm N}m_i/c^3$. In Eq.~(\ref{eq:Pbdot}) we have neglected the dependence on the eccentricity, since the \psr\ orbit is practically circular. The sensitivity of the WD companion has been neglected as well, since generally $s_{\rm c} \ll s_{\rm p}$ \citep[see e.g.][for details]{will18}. In principle, $\dot{P}_{\rm b}$ also has additional multipole contributions, dominated by the tensorial quadrupole \citep{de92b,will18}. They are all well below the current measurement precision (cf.\ Eqs.~\ref{eq:pbdot:gw}) and (\ref{eq:PbdotInt})) and therefore can be ignored in the following calculations.

While $|\zeta| \lesssim 10^{-5}$ due to Solar system experiments \citep{bit03}, one cannot a priori assume that $\zeta s_{\rm p}$ and $\zeta s_{\rm p}^2$ are small as well. In fact, within Damour--Esposito-Far{\`e}se gravity it has been found that $\zeta s_{\rm p}^2$ can be of order unity even if $\zeta \rightarrow 0$. In this regime of the so-called ``spontaneous scalarization'' the strong-field gravity of a neutron star can lead to very large sensitivities, $s_{\rm p}$.

For a given theory, i.e.\ a given $\omega(\phi)$, and a given equation of state for neutron star matter, one can calculate the sensitivity $s_{\rm p}$, and use the three post-Keplerian parameters, i.e.\ Eqs.~(\ref{eq:rShap}--\ref{eq:Pbdot}), with the two unknown masses as a consistency test \citep[see e.g.][]{dt92}. Here we will follow a more generic approach by leaving $\omega(\phi)$ unspecified and only making use of the Solar system constraint on $|\zeta|$. For a given $\zeta$, Eqs.~(\ref{eq:rShap}--\ref{eq:Pbdot}) can then be solved for $\tilde m_{\rm p}$, $\tilde m_{\rm c}$, and $s_{\rm p}$.

The numerical values for the Shapiro range in Eq.~(\ref{eq:rShap}) and the Shapiro shape~(\ref{eq:sShap}) can directly be taken from  Table~\ref{tab:param}: $r = 1.029 \pm 0.005 \, \mu{\rm s}$ and $s = \sin i = 0.998005 \pm 0.000065$. In Eq.~(\ref{eq:Pbdot}) we need the intrinsic change of the orbital period, i.e.\ we need to subtract Shklovskii ($\dot{P}_{\rm b}^{\rm Shk}$) and Galactic ($\dot{P}_{\rm b}^{\rm Gal}$) contributions from the observed orbital period derivative $\dot{P}_{\rm b}$ in Table~\ref{tab:param} \citep[see e.g.][]{lk05}. Using the parallax distance $d_\pi$ from Table~\ref{tab:param} and the Galactic potential of \cite{McMillan:2017} we find for the intrinsic change of the orbital period
\begin{equation}
   \dot{P}_{\rm b}^{\rm int} = \dot{P}_{\rm b}^{\rm obs} -
   \dot{P}_{\rm b}^{\rm Shk} - \dot{P}_{\rm b}^{\rm Gal}
   = -4.4_{-7.9}^{+7.7} \,{\rm fs\,s^{-1}}
   \label{eq:PbdotInt}
\end{equation}
It is worth mentioning that as discussed in Section~\ref{sssec:astro measure}, the observed secular change in orbital period is predominantly coming from the Shklovskii contribution. There is no measurable intrinsic orbital period change, as one can see from Eq.~(\ref{eq:PbdotInt}), and the Galactic correction is only $2.7 \pm 0.1 \,{\rm fs\,s^{-1}}$ and therefore smaller than the error in the Shklovskii correction. For comparison, with the masses from Table~\ref{tab:param} one finds for the orbital period change as predicted by the GR's quadrupole formula $ \dot{P}_{\rm b}^{\rm GR} = -2.79 \pm 0.03 \,{\rm fs\,s^{-1}}$ (see Eq.~(\ref{eq:pbdot:gw})), significantly smaller than the error in $\dot{P}_{\rm b}^{\rm int}$.

\begin{figure}
  \centering
  \includegraphics[width=8.3cm]{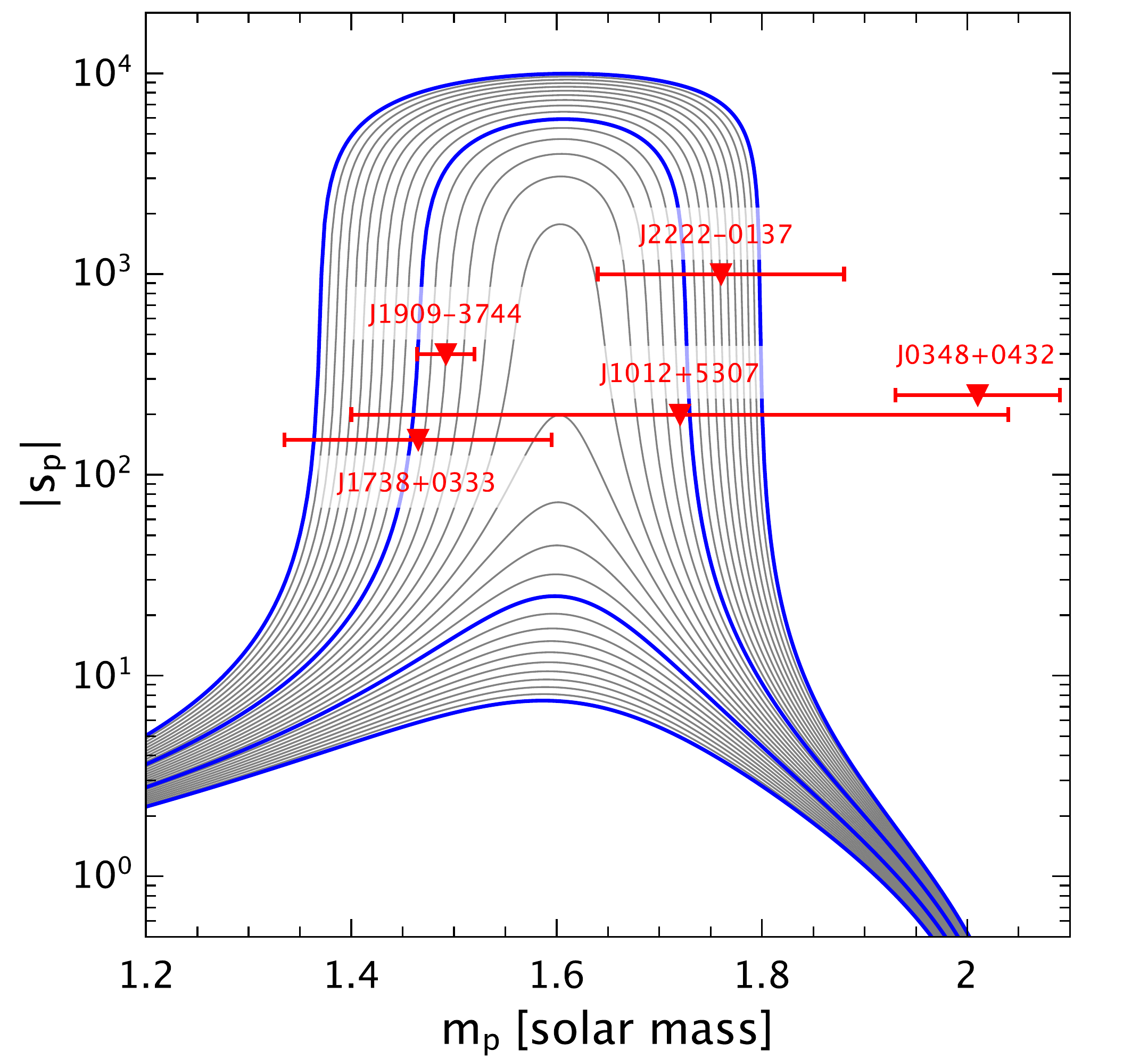}
  \caption{Upper limits on spontaneous scalarization within Damour--Esposito-Far{\`e}se (DEF) gravity from dipolar radiation tests with five different pulsar-WD systems, where $\zeta = 10^{-10}$ has been chosen. This figure is similar to Fig.~1 in \protect\cite{Shao:2017}, with the exceptions that the equation of state for neutron star matter is WFF1 \protect\citep{lp01}, the error bars for the masses are 2-$\sigma$, and our $y$-axis gives the absolute value of the pulsar's sensitivity $s_{\rm p}$ instead of the effective scalar coupling. Blue curves are plotted for different values of the coupling parameter $\beta$ in \protect\cite{de93} (from bottom to top: $\beta = -4.2,-4.3,-4.4,-4.5$). The thin grey lines indicate steps of 0.01 in $\beta$. Due to the precise mass measurement for \psr\, this pulsar provides the tightest constraint in that particular scenario, specifically excluding $\beta < -4.4$. Data for the other pulsars are taken from \protect\cite{afw+13} (J0348+0432), \protect\cite{ddf+20} (J1012+5307), \protect\cite{fwe+12} (J1738+0333), and \protect\cite{cfg+17} (J2222-0137).
  \label{fig:scalarization}}
\end{figure}

With the numbers for the three post-Keplerian parameters in hand, we can now use Eqs.~(\ref{eq:rShap}--\ref{eq:Pbdot}) to calculate the mass parameters $\tilde m_{\rm p}$ and $\tilde m_{\rm c}$ and put generic (i.e.\ independent of the details of $\omega(\phi)$) constraints on combinations of $\zeta$ and $s_{\rm p}$. As it turns out, for the allowed range of $|\zeta| \lesssim 10^{-5}$ the constraints on $s_{\rm p}$ do come exclusively from Eq.~(\ref{eq:Pbdot}). As a result, for all values of $\zeta$ compatible with Solar system experiments one has
\begin{equation}\label{eq:splimit}
    \sqrt{|\zeta|} \, |s_{\rm p}| < 4.0 \times 10^{-3} \quad \mbox{(95\% C.L.)} \,.
\end{equation}
In a comparison with the currently best limit on dipolar radiation obtained from PSR~J1738+0333 \citep{fwe+12,zdw+19} one has to keep in mind that $\zeta = \kappa_{\rm D}/4$ \citep{will18}. Assuming Eq.~(7) in \cite{zdw+19} for the sensitivity, in order to make the limit comparable, one obtains from Eq.~(\ref{eq:splimit})
\begin{equation}
  |\kappa_{\rm D}| < 2.0 \times 10^{-3} \quad \mbox{(95\% C.L.)} \,.
\end{equation}
This limit is still about an order of magnitude weaker than the one from PSR~J1738+0333. Nevertheless, the limit is qualitatively different as it is based on three post-Keplerian parameters, while the PSR~J1738+0333 limit used optical observations and WD models to obtain the masses of pulsar and companion. As a result, the mass of \psr\ is known with higher precision (better than 1\%; see Tab.~\ref{tab:param}), which is important for strong field effects that depend critically on the neutron-star mass. In Fig.~\ref{fig:scalarization} we give a specific example to illustrate this.

\subsubsection{Testing for a preferred frame for gravity}

High-precision timing with binary pulsars is a also powerful tool in constraining the gravitational Lorentz invariance violation via its orbital dynamics. This type of study is typically carried out using the parameterized post-Newtonian (PPN) formalism \citep{wil14, wex14, sw16, will18}. Here we investigate the application of the \psr{} system to such gravity experiments.

In the PPN framework, Lorentz invariance violation is described by three
PPN parameters, $\alpha_1$, $\alpha_2$, and $\alpha_3$. While $\alpha_1$
and $\alpha_2$ only pertain to semi-conservative dynamics, $\alpha_3$
additionally introduces nonconservative effects, namely, it simultaneously
introduces preferred-frame effects and violates the energy-momentum
conservation laws \citep{wil14, will18}. The $\alpha_3$ parameter was
severely bound, down to the level of $\sim 10^{-20}$, via the long-term
high-precision timing of PSR~J1713+0747 \citep{zdw+19}. Therefore, here we only consider constraints of $\alpha_1$ and $\alpha_2$. Because neutron
stars are strongly self-gravitating objects, in order to distinguish them
from weak-field objects, in the following we use $\hat\alpha_1$ and
$\hat\alpha_2$ respectively to denote the strong-field counterparts of
$\alpha_1$ and $\alpha_2$.

As shown in \citet{sw12}, for a nearly circular binary orbit,
$\hat\alpha_2$ introduces a precession of the orbital angular momentum
around the direction of ${\bf w}$, where ${\bf w}$ is the {\it absolute}
velocity of the binary with respect to a preferred frame. Usually, the
frame wherein the cosmic microwave background (CMB) is isotropic is chosen.
The $\hat\alpha_2$-induced precession causes the change of the orientation
of the orbit with respect to the observer, notably that the orbital
inclination angle $i$ is changing. This change introduces a nonzero time
derivative of the projected semi-major axis, and can be bound via the timing
parameter $\dot x$. The contribution is given by \citep{sw12},
\begin{align} \label{eq:xdot:a2}
  \left(\frac{\dot{x}}{x}\right)^{\hat{\alpha}_{2}}=-\frac{\hat{\alpha}_{2}}{4}
  n_{b}\left(\frac{w}{c}\right)^{2} \cot i \sin 2 \psi \cos \vartheta \,,
\end{align}
where $n_b \equiv 2\pi/P_b$, $\psi$ is the angle between ${\bf w}$ and the
orbital norm $\hat{\bf k}$, and $\vartheta$ is the angle between
$\mathbf{w}_{\perp} \equiv {\bf w} - \left( {\bf w} \cdot \hat{\bf k}
\right) \hat{\bf k}$ and the direction of ascending node. As discussed in Section~\ref{ssec:timing}, for PSR~J1909$-$3744 the proper motion effect is able to fully account for the observed $\dot x$. Hence, there is no need to have a nonzero $\hat\alpha_2$ in order to {\it
reproduce} the measured $\dot x$, which is in fact translated into a bound on $\hat\alpha_2$.

\begin{figure}
  \centering
  \includegraphics[width=8cm]{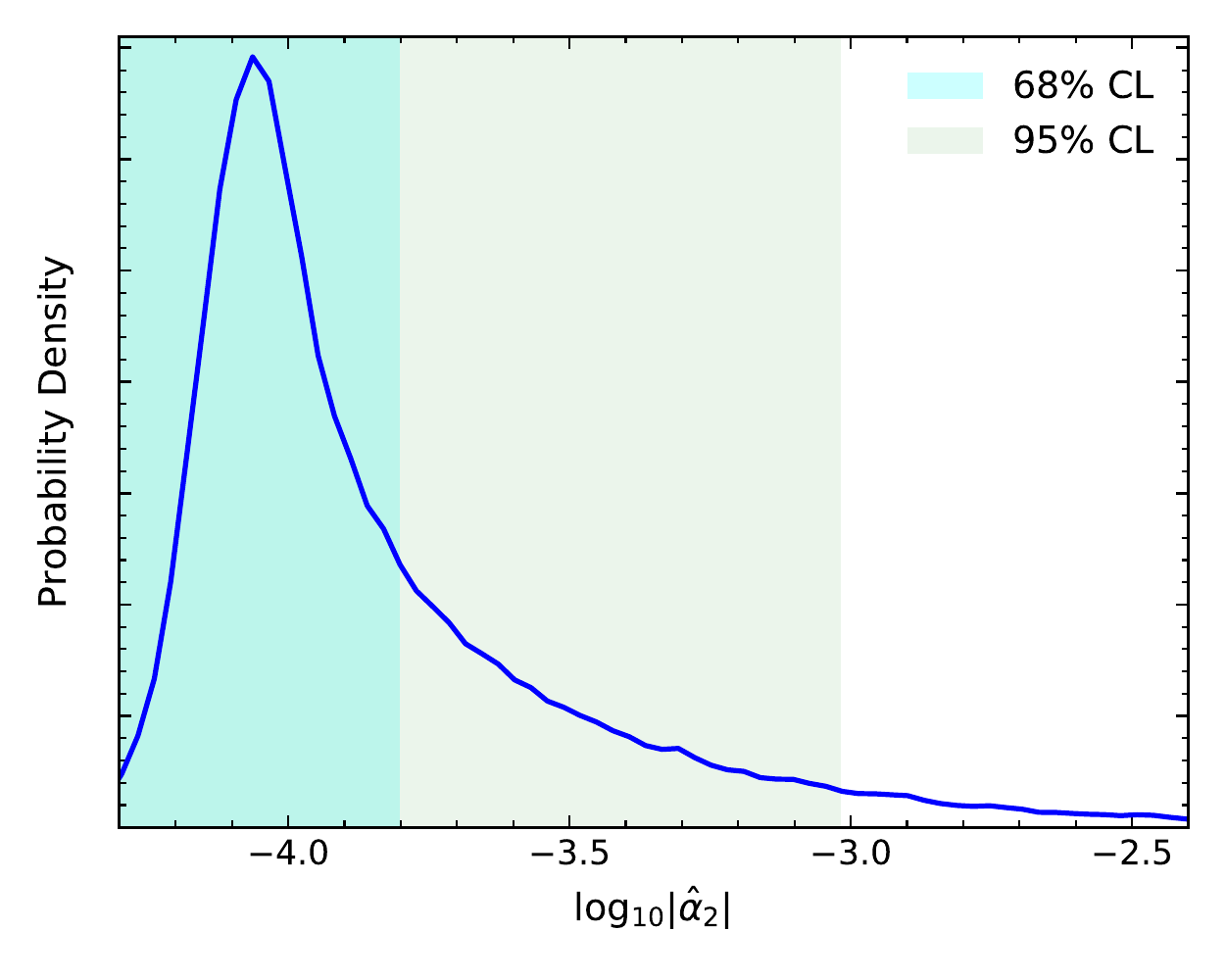}
  \caption{Probability distribution of $\hat\alpha_2$. The limits at 68\% and 95\%
confidence levels are shaded. Notice that the $\hat\alpha_2$ test here is a probabilistic test, as explained in detail in Section~3 of \citet{sw12}. \label{fig:a2}}
\end{figure}

We follow the method developed in \citet{sw12} to obtain the upper limit on $\hat\alpha_2$. We randomize orbital ascending node $\Omega$
uniformly in $\left[0, 360^\circ \right)$, and give equal probabilities to
the configuration $i \in \left[0, 90^\circ \right)$ and the configuration
$i \in \left[90, 180^\circ \right]$. We subtract the contribution to $\dot
x/x$ from the proper motion (\ref{eq:xdot:PM}) and assign the remaining
$\dot x/x$ to the $\hat\alpha_2$-induced precession (\ref{eq:xdot:a2}). Such a treatment renders the test a probabilistic one \citep[see Sec. 3 in][]{sw12}, thus the obtained results on $\hat\alpha_2$ should only be interpreted as upper bounds (see Fig.~\ref{fig:a2}). In
the test, the CMB frame is chosen to be the preferred frame, and the {\it
absolute} 3-dimensional systematic velocity is obtained for the binary by
combining radio timing and optical observations \citep{ant13}. All
uncertainties in relevant parameters are taken into account in the Monte Carlo
simulations. The probability density for $\hat\alpha_2$ is given in
Fig.~\ref{fig:a2}, from where we have,
\begin{align}
  \left| \hat\alpha_2 \right| &< 1.6 \times 10^{-4} \quad \left(\mbox{68\%
  C.L.}\right) \,, \\
  \left| \hat\alpha_2 \right| &< 9.7 \times 10^{-4} \quad \left(\mbox{95\%
  C.L.}\right) \,.
\end{align}
These limits are weaker than the best existing limit from the spin precession
of the Sun \citep{nor87} and solitary pulsars \citep{sck+13}, yet it still
represents a new modest limit from a completely different system.

As for the PPN parameter $\hat\alpha_1$, \citet{de92a} elegantly picturised
the {\it orbital polarization} phenomenon for near-circular binary orbits.
In this picture, a nonzero $\hat\alpha_1$ induces a polarization of the
eccentricity vector, ${\bf e} \equiv e \hat{\bf a}$ where $\hat{\bf a}$ is
the direction to the periastron. The {\it polarization} is towards a
direction in the orbital plane that is perpendicular to the {\it absolute}
velocity. Consequently, the observed eccentricity vector is a vectorial
superposition of a constant-length, {\it relativistically rotating}
eccentricity, ${\bf e}_R(t)$, and a {\it forced eccentricity}, ${\bf e}_F$,
\begin{align}
  \mathbf{e}(t)=\mathbf{e}_{F}+\mathbf{e}_{R}(t) \,.
\end{align}
The forced eccentricity is given by \citep{de92a, sw12},
\begin{align}\label{eq:eF}
  \mathbf{e}_{F}=\frac{\hat{\alpha}_{1}}{4 c^{2}} \frac{q-1}{q+1}
  \frac{n_{b}}{\dot{\omega}_{\mathrm{PN}}} \mathcal{V}_{O} \hat{\mathbf{k}}
  \times \mathbf{w} \,,
\end{align}
where $q \equiv m_p / m_c$, ${\cal V}_O \equiv \left({\cal G} M n_b
\right)^{1/3}$ with ${\cal G}$ the effective gravitational constant and $M
\equiv m_p + m_c$. The ${\dot{\omega}_{\mathrm{PN}}}$ in Eq.~(\ref{eq:eF})
is the constantly rotating rate of ${\bf e}_R(t)$, which, in GR, is the
well-known periastron advance rate.

The above picture was applied in \citet{sw12} to PSRs~J1012+5307 and
J1738+0333, and the yet tightest limit on $\hat\alpha_1$ was achieved. The
test used the observed eccentricity, decomposed into the Laplace parameters
$\eta \equiv e \sin\omega$ and $\kappa \equiv e \cos\omega$ \citep{lcw+01}.
It requires that the $\hat\alpha_1$-introduced time variations in $\eta$
and $\kappa$ to be consistent with their observed uncertainties
\citep{sw12}. The best constraint, $\hat{\alpha}_{1}=-0.4_{-3.1}^{+3.7}
\times 10^{-5}$ at the 95\% confidence level, comes from PSR~J1738+0333
\citep{fwe+12}.

To obtain constraint on $\hat\alpha_1$ with PSR~J1909$-$3744, we repeated the same timing analysis as described in Section~\ref{sssec:data}, with time derivatives of $\eta$ and $\kappa$ ($\dot\eta$ and $\dot\kappa$) directly included as fitted timing parameters. This allowed us to have high-precision measurements
of both $\eta$ and $\kappa$ and their time derivatives, and enabled us to perform a more straightforward test of $\hat\alpha_1$ by directly using the
time derivative of ${\bf e}(t)$. With a nonzero $\hat\alpha_1$, after
averaging over an orbital timescale, we have \citep{sw12}
\begin{align} \label{eq:edot}
  \dot{\mathbf{e}}(t)  = e
  \dot{\omega}_{\mathrm{PN}} \hat{\mathbf{b}}+\frac{\hat{\alpha}_{1}}{4
  c^{2}} \frac{q-1}{q+1} n_{b} \mathcal{V}_{O} \mathbf{w}_{\perp} \,,
\end{align}
where $\hat{\bf b} \equiv \hat{\bf k} \times \hat{\bf a}$, and
$\mathbf{w}_{\perp}$ is the projection of ${\bf w}$ onto the orbital plane.

\begin{figure}
  \centering
  \includegraphics[width=8.7cm]{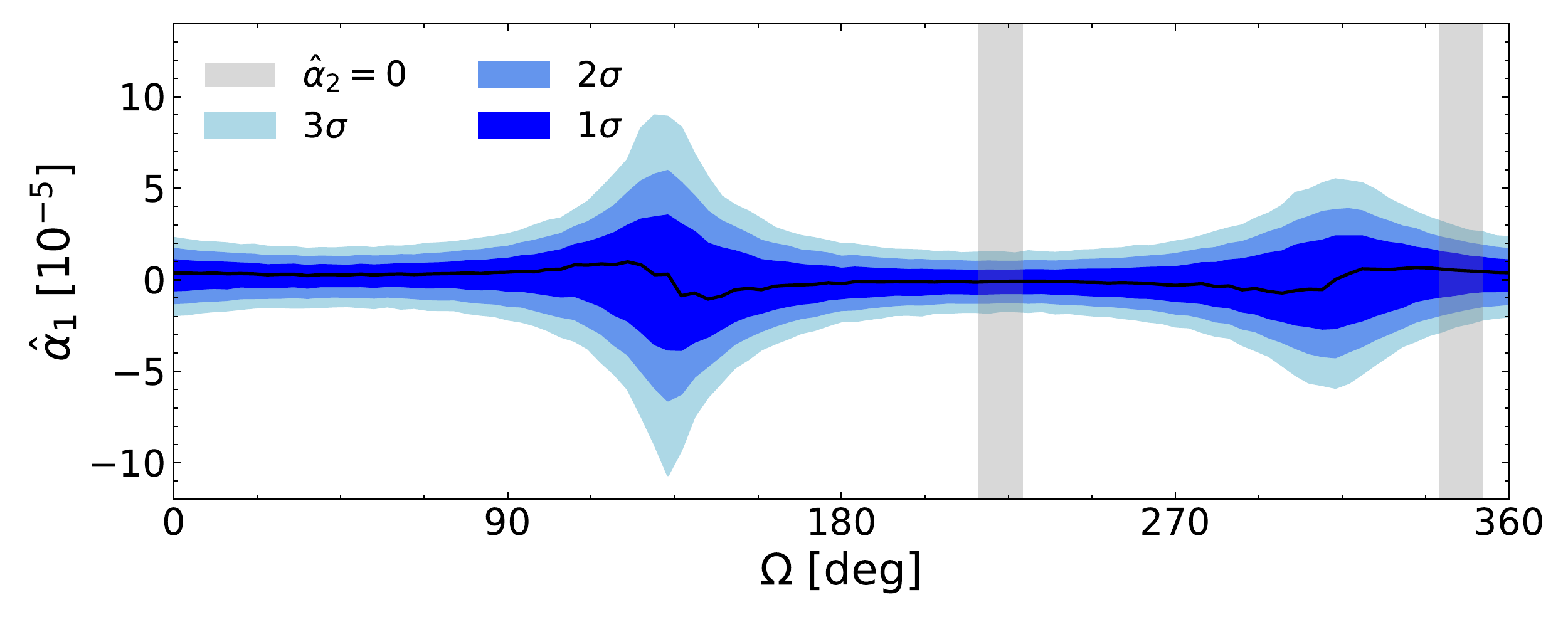}
  \caption{Constraint on $\hat\alpha_1$ as a function of the unknown longitude of the
ascending node, $\Omega$. The vertical gray bands give the allowed values
of $\Omega$ if we have assumed $\hat\alpha_2=0$ (cf. Fig.~\ref{fig:xdot}).
\label{fig:a1}}
\end{figure}

We set up Markov-chain Monte Carlo simulations which account for all the
uncertainties. For each value of $\Omega$, we use the {\sc emcee} package
\citep{fhlg13} to explore the posterior distribution of $\hat\alpha_1$ that
is consistent with both $\dot\eta$ and $\dot\kappa$. From the posterior, we
obtain the median, as well as the 1-$\sigma$, 2-$\sigma$, and 3-$\sigma$
enclosed regions for $\hat\alpha_1$. The result for the configuration $i <
90^\circ$ is given in Fig.~\ref{fig:a1}, while the other configuration with $i
\geq 90^\circ$ is merely a mirrored case of Fig.~\ref{fig:a1} \citep[see
e.g. Figs.~6 \& 7 in][]{sw12}. From the figure, we see that the loosest
limit is from $\Omega \sim 135^\circ$,
\begin{align}
  \left| \hat\alpha_1 \right| &< 3.7 \times 10^{-5} \quad \left(\mbox{68\%
  C.L.}\right) \,, \\
  \left| \hat\alpha_1 \right| &< 6.3 \times 10^{-5} \quad \left(\mbox{95\%
  C.L.}\right) \,.
\end{align}
The limit is slightly worse than that from PSR~J1738+0333.
If we assume that the $\hat\alpha_2$ limit from the spin precession of
solitary pulsars \citep{nor87, sck+13} is applicable to PSR~J1909$-$3744,
we can use Eq.~(\ref{eq:xdot:PM}) to determine $\Omega$ (cf.
Fig.~\ref{fig:xdot}). The corresponding allowed values of $\Omega$ are indicated by the gray bands in Fig.~\ref{fig:a1}, from which we can obtain,
\begin{align}
  \left| \hat\alpha_1 \right| &< 1.2 \times 10^{-5} \quad \left(\mbox{68\%
  C.L.}\right) \,, \\
  \left| \hat\alpha_1 \right| &< 2.1 \times 10^{-5} \quad \left(\mbox{95\%
  C.L.}\right) \,.
\end{align}
The limit is fractionally better than that from PSR~J1738+0333.
More importantly, our limit here is obtained via a more straightforward
method, directly using Eq.~(\ref{eq:edot}), and sophisticated statistical
analysis was carried out using the Bayesian inference.


\section{Conclusions} \label{sec:conclu}

In this paper, we presented a high-precision timing analysis and an astrophysical study of the binary millisecond pulsar, PSR~J1909$-$3744. We have managed to achieve a timing precision of approximately 100\,ns with 15 years of data collected with the Nan\c{c}ay Radio Telescope. The measurements out of the timing analysis have been examined by using broad-band and sub-band TOA datasets, by incorporating jitter noise in the TOA errors and by conducting the analysis with different software (\textsc{temponest} and \textsc{enterprise}). We have improved measurement precision of secular changes in orbital period and projected semi-major axis, and showed that these measured variations are dominated by the relative motion between the pulsar system and the barycenter. Using the orbital period derivative measurement, we derived a kinematic distance of the system which is highly consistent and more precise than the parallax distance. In addition, we identified four possible solutions to the ascending node of the pulsar orbit with the secular change in the projected semi-major axis.

By combining our timing measurements and published observations of the WD companion, we investigated the binary evolution history of the PSR~J1909$-$3744 system by modelling with the stellar evolution code MESA, and discussed solutions based on detailed WD cooling at the edge of the WD age dichotomy paradigm. We additionally determined the 3-D velocity of the system and depicted its highly eccentric orbit around the centre of our Galaxy. Moreover, we placed a constraint over dipolar gravitational radiation, a complement to previous studies given the precisely measured mass of the pulsar. We derived an improved limit on the (strong-field counterpart of) PPN parameter, $\hat\alpha_1<2.1 \times 10^{-5}$ at 95\% confidence level, achieved with a more concrete method than previous analysis.


\section*{Acknowledgements}
We are grateful to V.~Venkatraman~Krishnan for valuable discussions. We also would like to thank the anonymous referee for providing very constructive comments that have helped to improve the manuscript. KL, GD and NW are supported by the European Research Council for the ERC Synergy Grant BlackHoleCam under contract no.\ 610058. LS was supported by the National Natural Science Foundation of China (11975027, 11991053, 11721303), the Young Elite Scientists Sponsorship Program by the China Association for Science and Technology (2018QNRC001), and the Max Planck Partner Group Program funded by the Max Planck Society.  AGI acknowledges support from the Netherlands
Organisation for Scientific Research (NWO) and is thankful for very productive discussions with Gijs Nelemans, Alejandra Romero, Gabriel Lauffer and Pablo Marchant.
The Nan\c{c}ay Radio Observatory is operated by the Paris Observatory, associated with the French Centre National de la Recherche Scientifique. We acknowledge financial support  from the Action F\'ed\'eratrice PhyFOG funded by Paris Observatory and from “Programme National Gravitation, R\'ef\'erences, Astronomie, M\'etrologie” (PNGRAM) funded by CNRS/INSU-IN2P3-INP, CEA and CNES, France.

\section*{Data Availability}

The timing data used in this article shall be shared on reasonable request to the corresponding author.

\bibliographystyle{mnras}
\bibliography{journals,psrrefs,modrefs,crossrefs}

\label{lastpage}
\end{document}